# MACHINE LEARNING CHARACTERIZATION OF CANCER PATIENTS-DERIVED EXTRACELLULAR VESICLES USING VIBRATIONAL SPECTROSCOPIES: RESULTS FROM A PILOT STUDY.

[1]Abicumaran Uthamacumaran, [2]Samir Elouatik, [3,4]Mohamed Abdouh, [5]Michael Berteau-Rainville, [6]Zu-hua Gao, and [7,8,9]*Goffredo Arena

[1]*Concordia University, Department of Physics, Montreal, QC, Canada*

[2] *Université de Montréal, Département de chimie, Montreal, QC, Canada*

[3] Cancer Research Program, Research Institute of the McGill University Health Centre, 1001 Decarie Boulevard, Montreal, Quebec, Canada, H4A 3J1

[4] The Henry C. Witelson Ocular Pathology Laboratory, McGill University, Montreal, QC, Canada

[5]*Institut national de la recherche scientifique, Centre Énergie Matériaux Télécommunications, Varennes, QC, Canada,* J3X 1P7

[6] Department of Pathology and Laboratory Medicine, University of British Columbia, G105-2211 Wesbrook Mall, Vancouver BC, Canada V6R 2B5

[7] Istituto Mediterraneo di Oncologia, Viagrande, Italy

[8] Department of Surgery, McGill University, St. Mary Hospital, 3830 Lacombe Avenue, Montreal, Quebec, Canada, H3T 1M5

[9] Fondazione Gemelli-Giglio, Contrada Pollastra, Cefalu'

*Correspondence:* Goffredo Arena* (Primary Supervisor): goffredoarena@gmail.com

# ABSTRACT

**Background:** The early detection of cancer is a challenging problem in medicine. The blood sera of cancer patients are enriched with heterogeneous secretory lipid-bound extracellular vesicles (EVs), which present a complex repertoire of information and biomarkers, representing their cell of origin, that are being currently studied in the field of liquid biopsy and cancer screening. Vibrational spectroscopies provide non- invasive approaches for the assessment of structural and biophysical properties in complex biological samples.
**Methods:** In this pilot study, multiple Raman spectroscopy measurements were performed on the EVs extracted from the blood sera of n= 9 patients consisting of four different cancer subtypes (colorectal cancer, hepatocellular carcinoma, breast cancer and pancreatic cancer) and five healthy patients (controls). FTIR (Fourier Transform Infrared) spectroscopy measurements were performed as a complementary approach to Raman analysis, on two of the four cancer subtypes. The spectra were subjected to various machine learning classifiers with hyperparameter optimization to discriminate between healthy and cancer patients-derived EVs.

**Results & Applications:** The AdaBoost Random Forest Classifier, Decision Trees, and Support Vector Machines (SVM) distinguished the baseline corrected Raman spectra of cancer EVs from those of healthy controls (N=18 spectra) with a classification accuracy of > 90 % when reduced to a spectral frequency range of $1800 - 1940$ $cm^{-1}$ and subjected to a 50:50 training: testing split. FTIR classification accuracy on N=14 spectra showed an 80% classification accuracy. Our findings demonstrate that basic machine learning algorithms are powerful applied intelligence tools to distinguish the complex vibrational spectra of cancer patient EVs from those of healthy patients. These experimental methods hold promise as valid and efficient liquid biopsy for artificial intelligence-assisted early cancer screening.

## 1. INTRODUCTION

Cancers remain globally one of the leading causes of disease-related death, accounting for nearly 10 million deaths in 2020 (World Health Organization, 2021). Early diagnosis, using different strategies, remains the best way to decrease mortality from neoplastic disease. However, the early detection of cancer remains a challenging real-life complex problem requiring interdisciplinary systems thinking and applied intelligence. The sera of cancer patients are enriched with extracellular vesicles (EVs) (Théry et al., 2018). EVs are nanoscopic (~30-200 nm), heterogeneous packets of information released by cells forming long-range, intercellular communication networks (Uthamacumaran, 2020; Théry, et al., 2018). EVs are central mediators and control mechanisms of cancer cybernetics. Complex adaptive behaviors such as therapy resistance and phenotypic transitions within cancer ecosystems are in large part conferred by EV flows (Samuel et al., 2017; Ramakrishnan et al., 2020; Fontana et al., 2021). EVs are emergent, reprogramming machineries used by cancer ecosystems to regulate cell fate dynamics (Guo et al., 2019). Furthermore, EVs can reprogram distant tissue microenvironments into pre-metastatic niches and horizontally transfer malignant traits to target cells located in distant organs (Abdouh et al., 2014; 2016; 2017; 2019; 2020; Arena et al., 2017; Hamam et al.,2016, Guo et al., 2019) as well as therapy resistance to promote aggressive cancer phenotypes (Steinbichler et al., 2019; Keklikoglou et al., 2019). On the other hand, EVs-derived from an hESC (human embryonic stem cell) microenvironment have been shown to suppress cancer phenotypes and reprogram a subset of malignant phenotypes to more benign states (Camussi et al., 2011; Zhou et al., 2017).

EVs are currently being studied as potential biomarkers in liquid biopsy for cancer screening (Zhao et al.,2019). However, the ability to distinguish cancer patients from healthy patients according to EVs patterns remains challenge due to the lack of known specific EVs cancer markers. Raman spectroscopy provides a non-invasive experimental approach for the assessment of structural and biophysical properties in complex systems (Ember et al., 2017). The Raman technique depends on the change in polarizability of a molecule (due to the inelastic scattering of light by molecules) and measures the relative frequencies at which scattering of radiation occurs, while IR spectroscopy depends on a change in the dipole moment (Smith and Dent, 2005; Larkin, 2011).
The general principles of Raman scattering can be summarized as follows. When a photon interacts with a molecule, it provides energy to induce a short-lived energy transition of the vibrational state of the molecule towards an unstable, higher-energy transition state. We call this Raman scattering. In contrast, the IR spectra would provide information about Raman inactive vibrational modes better suited for asymmetric polar mode vibrations. Conventionally, vibrational spectroscopy techniques have been used to elucidate the molecular structures of biochemical samples via Raman peak assignment (Brusatori et al., 2017). In the present project we applied Raman and FTIR spectroscopy to human sera EVs to evaluate for the presence of robust patterns able to characterize cancer EVs from non-cancer EVs in liquid biopsy for cancer screening.

The EVs in the patient-derived liquid biopsies consist of a complex mixture of heterogeneous EVs with a diverse set of information. To overcome the challenge represented by the potential inadequacy of these methods to capture the spectral complexity of cancer-derived EVs, we implemented the use of artificial intelligence (AI) to help in the detection of patterns in complex systems such as cancer EVs. Statistical Machine Learning (ML) algorithms, a subset of AI, are state-of-the-art approaches for pattern detection in complex real-life problems through data science (Uthamacumaran, 2020) and given the complexity of these systems, we hypothesized that the use of ML algorithms might provide a more efficient way of detecting characteristic patterns and classifying EVs between healthy and cancer group. In our pilot study, we demonstrate for the first time the use of Random Forest, Decision Trees, and Support Vector Machines as ML classifiers in accurately discriminating healthy patient-derived EVs vibrational spectra from those of cancer patients.

## 2. MATERIALS AND METHODS.

### 2.1 EVs EXTRACTION, PURIFICATION, AND CHARACTERIZATION.

#### 2.1.1. Blood collection and serum preparation:

Patients for the current study were recruited form the department of General Surgery at the Royal Victoria Hospital and St-Mary's Hospital (Montreal, Canada) and underwent a written and informed consent for blood collection in accordance to a protocol approved by the Ethics Committee of our institution (SDR-10-057). Blood samples were collected from both healthy individuals and patients who presented to our clinic for a follow-up or those that underwent resection of primary cancer (Table 1). Blood samples (10 ml) were collected from a peripheral vein in vacutainer tubes (Becton Dickinson) containing clot-activation additive and a barrier gel to isolate serum. Blood samples were incubated for 60 min at room temperature to allow clotting and were subsequently centrifuged at 1500 g for 15 min. The serum was collected, and a second centrifugation was performed on the serum at 2000 g for 10 min, to clear it from any contaminating cells. Serum samples were aliquoted and stored at −80°C until use.

**TABLE 1.** Patient cases identification.

| Patient ID | Disease |
|---|---|
| Case 200717 | Breast cancer |
| Case 160517 | Pancreatic cancer |
| Case 426 | Colorectal cancer |
| Case 549 | Hepatocellular carcinoma |
| Case 161216 | Healthy control |
| Case 220916 | Healthy control |
| Case270717.1 | Healthy control |
| Case270717.2 | Healthy control |
| Case270717.3 | Healthy control |

2.1.2. EVs isolation on Iodixanol gradient:

Serum samples were diluted in cold phosphate buffered solution (PBS) and were subjected to a series of sequential differential centrifugation steps (Abdouh et al., 2019). Samples were centrifuged at 500 g for 10 min to remove contaminating cells, followed by centrifugation at 2000 g for 20 min to remove cell debris. Supernatants were passed through a 0.2 μm syringe filter (Corning) and centrifuged at 16,500 g for 20 min at 4ºC to remove apoptotic bodies and remaining cell debris. Supernatants were ultracentrifuged at 120,000 g (40,000 rpm) for 80 min at 4ºC using 70 Ti rotor in an Optima XE ultracentrifuge machine (Beckman Coulter). The pellets were recovered in PBS/120 mM (4-(2- hydroxyethyl)-1-piperazineethanesulfonic acid) (HEPES) buffer solution. EVs samples were purified using iodixanol (OptiPrep density gradient, Sigma). A density gradient was prepared by serial dilutions of iodixanol stock (60% w/v): (i) 5 volumes of 60% iodixanol were mixed with 1 volume of 0.25 M Sucrose, 0.9 M NaCl and 120 mM HEPES solution (SNH solution, pH 7.4) to obtain 50% iodixanol, (ii) 2 volumes of 50% iodixanol were mixed with 3 volumes of SNH solution to obtain 20% iodixanol, and (iii) 1 volume of 50% iodixanol was mixed with 9 volumes of SNH solution to prepare 5% iodixanol. EVs samples (1.92 ml) were mixed with 2.88 ml of 50% iodixanol to obtain a 30% iodixanol solution. The discontinuous iodixanol density gradient was prepared by carefully layering 2.5 ml of 5% iodixanol and 2.5 ml of 20% of iodixanol sequentially in 13.26 ml Ultra-Clear tubes (Beckman Coulter). EVs in 30% iodixanol solution (4.8 ml) were carefully placed at the bottom of the tubes without disturbing the gradient. Tubes were centrifuged at 120,000 g (40,000 rpm) for 2 hrs at 4ºC using SW41 Ti swinging bucket rotor. Nine fractions of one milliliter each were collected, diluted in PBS and centrifuged at 120,000 g (40,000 rpm) for 80 minutes. The crude EVs pellets were washed twice with PBS at 120,000g (40,000 rpm) for 80 min at 4ºC, resuspended in 200 μl PBS, and stored at -80ºC until use. For subsequent analyses, fractions 3-5 were pooled as they were enriched for EVs (See Figure 5 in Appendix).

### 2.1.3. EVs characterisation:

EVs were characterized physically using Transmission Electron Microscopy (TEM) and NanoSight Nanoparticle Tracking Analysis (NTA), and phenotypically via Western blot. TEM analyses were performed using a JEM-2010 electron microscope (Jeol Ltd.). 20 µl of EVs samples were loaded on a copper grid and stained with 2% phosphotungstic acid. Samples were dried by incubating them for 10min under an electric incandescent lamp and scanned.

In Parallel, diluted aliquots of the extracted EVs samples were run on a Nanosight NS500 system (Nanosight Ltd.), and the size distribution was analyzed using the NTA 1.3 software. For immunoblotting, pelleted EVs were lysed in (Radioimmunoprecipitation Assay) (RIPA) buffer containing protease inhibitors (Sigma). Equal amounts of proteins were resolved on 12% sodium dodecyl sulphate– polyacrylamide gel electrophoresis (SDS-PAGE) and transferred to a nitrocellulose membrane (BioRad). Membranes were blocked in tris-buffered saline (TBS) containing 5% non-fat drymilk and exposed overnight at 4°C to mouse-anti-TSG101 (ab83, Abcam), and mouse-anti-Alix (2127, Cell Signaling) antibodies. Membranes were washed 3 times in Tris-buffered saline with 0.05% Tween-20 (TBST) and incubated with anti-mouse horseradish peroxidase (HRP)-conjugated secondary antibody for 1 hr at room temperature. After 3 additional washes in TBST, the blots were developed using Immobilon Western HRP Substrate (Millipore, Etobicoke, Canada) (Théry et al., 2018).

### 2.2 RAMAN SPECTROSCOPY PROTOCOL

Thawed EVs samples prior to Raman measurements were lightly vortexed to resuspend the EVs in the buffer. Local Raman shifts in the samples were measured under ambient conditions using a Renishaw inVia micro-Raman confocal microscope system with a 514 nm excitation green laser. The samples were illuminated using a 50 X objective (numerical aperture of 0.75) and Raman spectra were recorded and analyzed using the Renishaw Wire software. Typical measurements began with the calibration of the instrument with a silicon wafer and background noise determination on a clean Raman-grade calcium fluoride ($CaF_2$) slide (Crystran Ltd, UK) at 10% laser power, subsequently followed by measurements of the PBS buffer. The target irradiance on the sample is 2.5 mW/µm² to avoid sample degradation. Then, multiple measurements of the patient- derived EVs samples were performed under identical conditions after the 20 µL droplets of the EVs suspension has been air-dried on the $CaF_2$ slide. Measurements were performed for 30 second exposure time and 10 accumulations to optimize the signal-to-noise ratio (S/N).

### 2.3 FTIR SPECTROSCOPY PROTOCOL

The FTIR measurements were performed in transmission mode with an Agilent FTIR microscope equipped with a (deuterated triglycine sulfate) DTGS detector. The samples were prepared as dried films of EVs/buffer mixture on $CaF_2$ windows (identical conditions as the Raman measurements). FTIR and Raman samples were prepared from the same EVs samples. The source power and magnification(15x) are standard and are fixed by the system configuration. The FTIR spectrum acquisition were performed with a 4 $cm^{-1}$ spectral resolution and 256 co-added scans.

### 3.4. SOFTWARE AND STATISTICAL ANALYSIS

Data pre-processing consisted mainly of baseline correction. The infrared spectra were baseline- corrected using the asymmetric least-squares smoothing method as implemented in OriginPro 2021. Forthe FTIR spectra, the asymmetric factor was set to $10^{-3}$, the threshold at $10^{-3}$ and the smoothing factor at 3. For the Raman spectra, the asymmetric factor was set at $10^{-4}$, the threshold at $10^{-3}$ and the smoothing factor at 3. Peak fitting was performed post-baseline correction (See Appendix). Machine Learning analysis was performed on Google Colab using the machine learning binary classifiers from the Scikit-learn python library (Pedregosa et al., 2011). The random forest algorithm builds and combines multiple decision trees for a greater accuracy in prediction of patterns in complex datasets. In decision trees, each tree predictor depends on the values of a random vector sampled independently (Breiman, 2001; Bishop, 2006). The random forest (RF) ensemble methods minimize the errors and optimize the variance-bias trade off in the datasets to provide a more accurate prediction in the data classification (Pedregosa et al., 2011; Geron, 2019). Herein, the AdaBoost classifier was used as an ensemble learner to enhance the predictive performance of the RF classifier. The hyperparameters of the three classifiers were tuned as follows:

*RF AdaBoost Classifier:* max_Depth = 6, min_samples_leaf = 3, min_samples_split = 10, n_estimators = 50, C= 1.0 (learning rate), and the SAMME.R algorithm.

*Decision Trees*: ccp_alpha=0.0, class_weight=None, criterion='entropy', max_depth=None, max_features=None, max_leaf_nodes=None, min_impurity_decrease=0.0, min_impurity_split=None, min_samples_leaf=1, min_samples_split=2, min_weight_fraction_leaf=0.0, presort='deprecated', random_state=None, splitter='best'.

*SVM*: C=1.0, break_ties=False, cache_size=200, class_weight=None, coef0=0.0, decision_function_shape='ovr', degree=3, gamma='scale', kernel='linear', max_iter=-1, probability=False, random_state=None, shrinking=True, tol=0.001, verbose=False.

## 3. RESULTS

### 3.1. Vibrational Spectra highlight characteristic peaks in patient-derived EVs.

The protein bands correspond to the peaks at 1440 cm$^{-1}$ (CH$_2$, CH$_3$ deformation), 1580 cm$^{-1}$ (amide II bond) and amide I protein band (1600–1690 cm$^{-1}$) visible in Figure 1. They were present in both healthy and cancer samples with no distinct observable patterns distinguishing the two. The lipid- bands (2750–3040 cm$^{-1}$) are clearly distinct in the baseline corrected Raman spectra and have a stronger intensity in some of the cancer EVs samples (except for Case200717) in comparison to the healthy samples (the EVs membrane is composed of lipids) (Figure 1A-D). The spectra show that the differences in vibrational modes are difficult to generalize amidst the different cancer groups using traditional peak fit assignment.

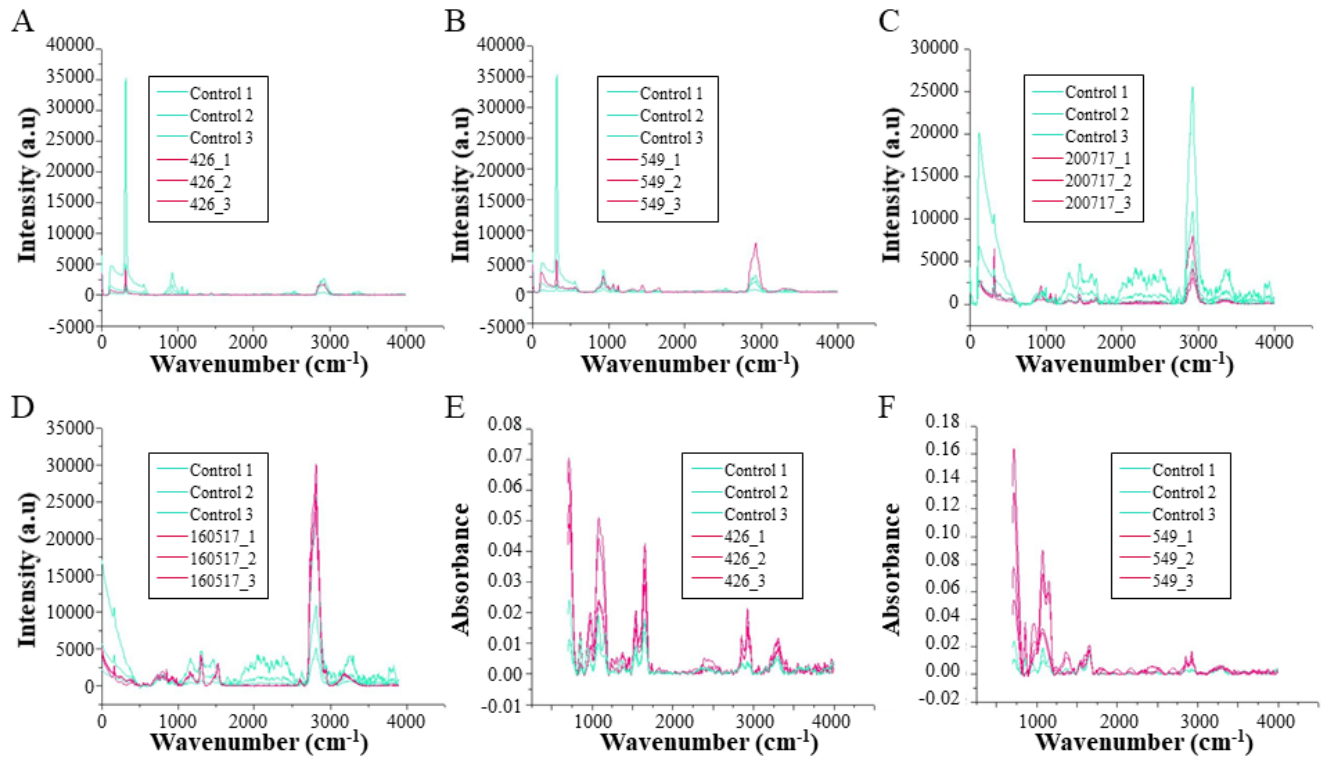

**FIGURE 1. BASELINE-CORRECTED EVs SPECTRA.** Raman (A-D) and FTIR (E-F) spectra of EVs isolated from health and cancer patient sera. **(A) BASELINE-CORRECTED RAMAN SPECTRA OF Case 426 (COLORECTAL CANCER).** The Raman spectra of cancer sample 426 (in pink) and healthy controls (in green) are shown as measured at a 514 nm excitation laser with 10% laser power, 10 accumulations for 30 seconds. Identical conditions were used for the following Raman spectra in the Graph 1 panel. EVs drops were air dried on the calcium fluoride slides prior to measurement acquisitions. The single peak at 321 cm$^{-1}$ corresponds to the CaF$_2$ substrate. **(B) BASELINE-CORRECTED RAMAN SPECTRA OF 549 (HEPATOCELLULAR CARCINOMA). (C) BASELINE - CORRECTED RAMAN SPECTRA OF CANCER Case 200717 (BREAST CANCER).** The single high intensity peak in the 2800-3010 cm$^{-1}$ corresponds to the lipid band. **(D) BASELINE-CORRECTED RAMAN SPECTRA OF CANCER Case 160517 (PANCREATIC CANCER). (E) BASELINE-CORRECTED FTIR SPECTRA OF 426**. **(F) BASELINE-CORRECTED FTIR SPECTRA OF 549**.

## 3.2. PCA (Principal Component Analysis) Dimensionality reduction accurately separates clusters of healthy patients' (control) EVs from cancer patients' EVs.

The PCA 2D scatter plot shows a pattern with data dispersion and lower cluster resolution along with a significant overlapping between the domains of normal population and cancer population data points (Figure 2). On the PCA plot, we can see a separation of different data domains. In analysis of the Raman spectra shown above, it was deduced that the spectral range from 1800-1940 cm$^{-1}$ would provide a better PCA clustering separation amidst the two groups (corresponding to Figure 2B and 4D). PCA clustering seems to well separate healthy FTIR spectra from cancer FTIR spectra, as well (Figure 2E and 4F). Our study shows that PCA, a multivariate linear dimensionality reduction, may be an efficient unsupervised, linear feature selection and is sufficient for the classification of patient-derived EVs. With an increased patient size, image classification Deep Learning networks such as convolutional neural networks (CNN) or Residual neural networks like ResNet can be trained with the PCA clustering of patient EV vibrational spectra as an additional layer of input in prospective studies.

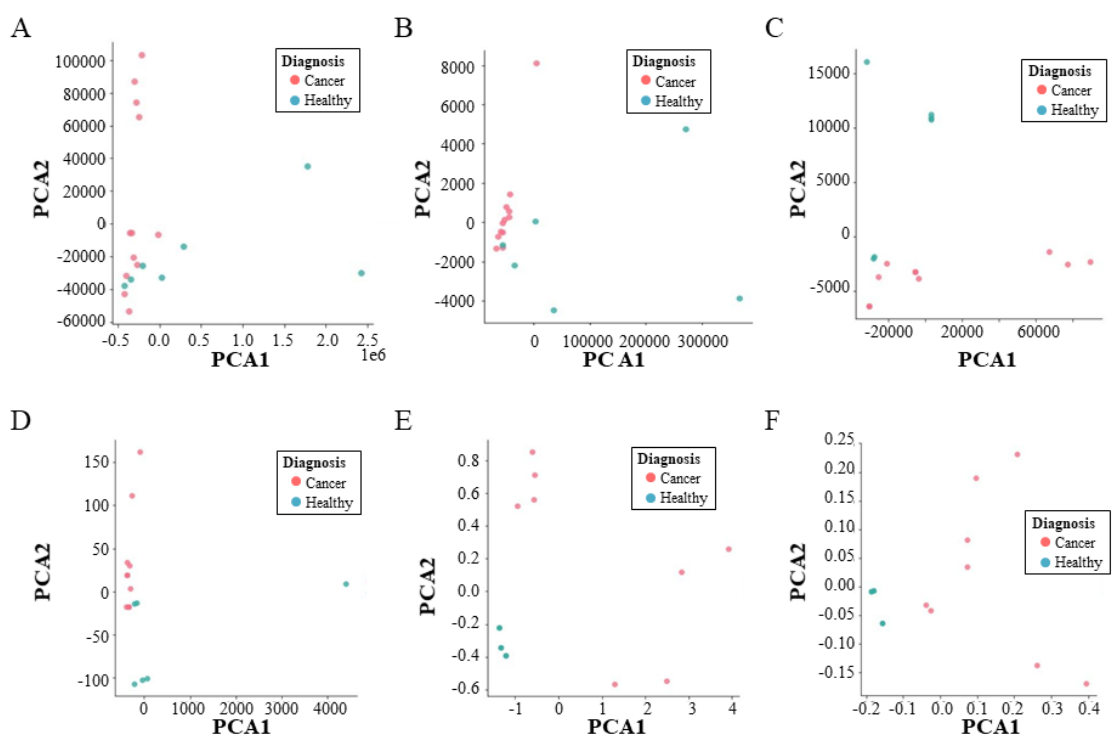

**FIGURE 2. PCA CLUSTERING ON EVs SPECTRA. (A). PCA Clustering on Raman Spectra without Baseline Correction (N=19).** The raw Raman spectra of 19 spectra (12 cancers and 7 controls) from n= 9 patients have been clustered by their first two principal components. Some of the spectral PCA points are overlapping between the cancer and healthy groups (in pink and green, respectively). (B) **PCA on Frequency Reduced Raman Spectra without Baseline Correction (N=18)**. (C) **PCA Clustering on Baseline- Corrected Raman Spectra (N=18).** The complete range Raman spectra when subjected to the baseline correction, shows better PCA separation amidst the binary classes. (D) **PCA Clustering on Baseline-Corrected Raman Spectra with Reduced Frequency (N=18) (**1800-1940 cm$^{-1}$). (E) **PCA Clustering on FTIR Spectra (N=14) without Baseline Correction.** The PCA space of raw FTIR spectra of 8 cancer samples (four of patient sample 426 and four of patient sample 549) and 6 healthy controls shows well- separated clusters even without baseline correction. (F) **PCA Analysis on Baseline- Corrected FTIR Spectra (N=14).**

### 3.3. ML algorithms show near 90% classification accuracy in distinguishing healthy and cancer EVs on Baseline-corrected Raman spectra in the reduced frequency range (1800-1940 cm$^{-1}$)

All tested binary classifiers well distinguished cancer EVs from healthy EVs with a near 90% classification accuracy on the reduced frequency Raman spectra with a 5-fold cross-validation (CV) (Figures 3A-H). Although the SVM algorithm showed the highest classification accuracy on the baseline-corrected Raman spectra within the reduced frequency range, the five-fold cross-validation score showed significant uncertainty (standard deviation) (Figures 3 G-H). These findings confirm that ML algorithms and hence, AI, can accurately distinguish between cancer and healthy patients- derived EVs on baseline corrected Raman spectra with high sensitivity and specificity.

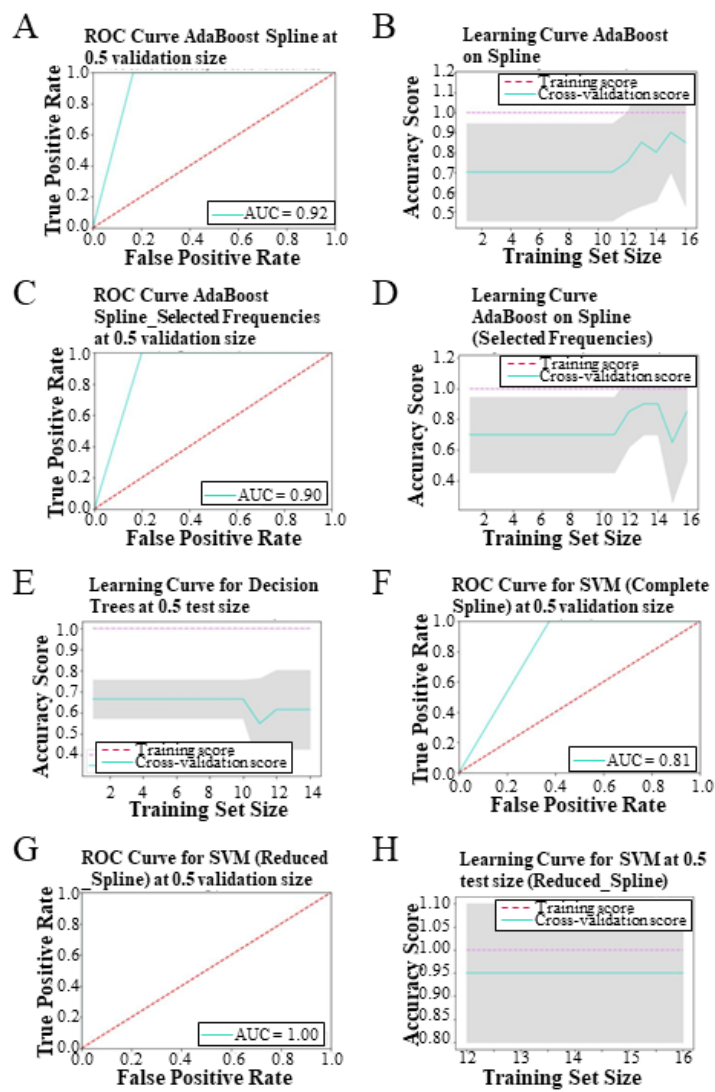

**FIGURE 3. ML PERFORMANCE ON BASELINE-CORRECTED RAMAN SPECTRA WITH 5-FOLD CROSS-VALIDATION (CV). (A)** AdaBoost Classifier's ROC on Baseline Corrected Raman Spectra (Full Range). **(B) ADABOOST CLASSIFIER'S CV LEARNING CURVE ON BASELINE CORRECTED RAMAN SPECTRA.** CV accuracy score: 90.00 ± 20.00 %. **(C) AdaBoost Classifier's ROC on Baseline Corrected Raman Spectra (Reduced Frequency Range). (D) ADABOOST CLASSIFIER'S CV LEARNING CURVE ON BASELINE CORRECTED RAMAN SPECTRA (REDUCED FREQUENCY).** CV accuracy score: 83.33 ± 21.08 %. **(E) CV Curve for Decision Tree on Baseline Corrected Raman Spectra.** CV accuracy score: 30.00 ± 24.45 %. **(F) ROC CURVE FOR SVM ON BASELINE CORRECTED RAMAN SPECTRA (FULL RANGE).** Classification accuracy of 66.66% and MSE: 0.333. The f1 scores of the cancer and healthy group classification were 0.77 (i.e., precision of 1.00 and recall of 0.62) and 0.40 (i.e., precision of 0.25 andrecall of 1.00), respectively corresponding to an AUC of 0.81. The CV accuracy score was found to be 80.00 ± 40.00 %. **(G) ROC CURVE FOR SVM PERFORMANCE ON BASELINE CORRECTED RAMAN SPECTRA WITH REDUCED FREQUENCY RANGE.** Classification accuracy of 100% and MSE: 0.0. The turquoise line is not visible and farthest away from the red-dashed line of the ROC curve indicating perfect classification accuracy. The f1 scores were 1.00 for both groups indicating a 100% sensitivity and specificity. **(H) CV CURVE FOR SVM PERFORMANCE ON REDUCED FREQUENCY RAMAN SPECTRA.** CV accuracy score: 90.00 ± 20.00 %

**3.4. Tree-based classification algorithms well-distinguish cancer FTIR spectra from healthy FTIR spectra with 90% classification accuracy.** Both tree-based machine learning algorithms, the AdaBoost RF and Decision Tree classifiers exhibited a decent classification accuracy in distinguishing cancer FTIR spectra from healthy patient FTIR spectra with an area under the curve of 0.83 and 1.00, respectively (Figures 4A-D). However, the SVM was found to be a poor classifier of the two patient groups' FTIR spectra (Figures 4E-F). These preliminary results demonstrate that Raman spectra may provide a more robust hyperparameter/model-dependent approach to characterizing and distinguishing cancer EVs from healthy EVs in comparison to FTIR.

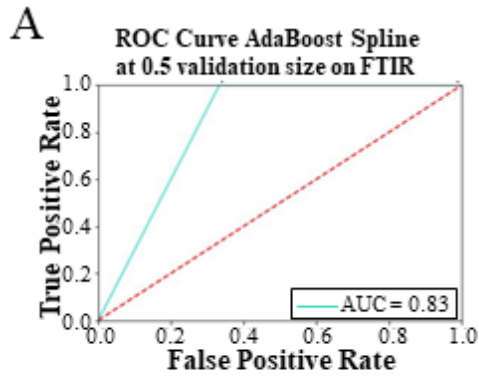 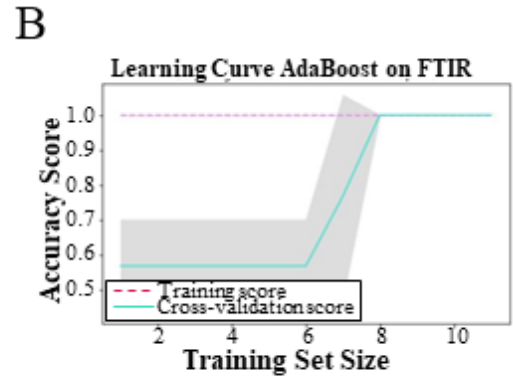
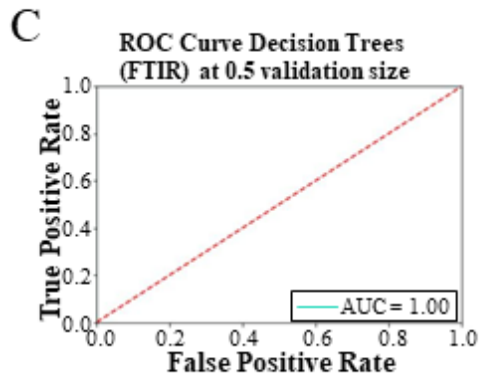 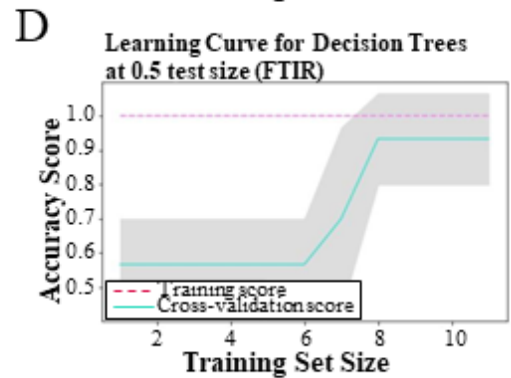
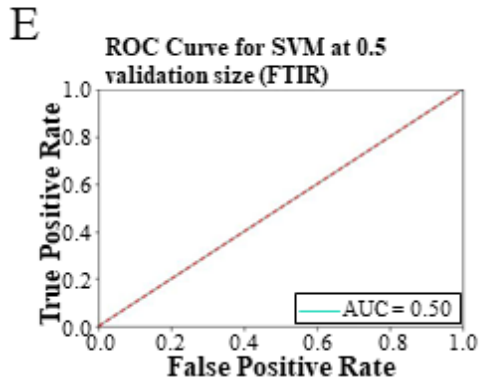 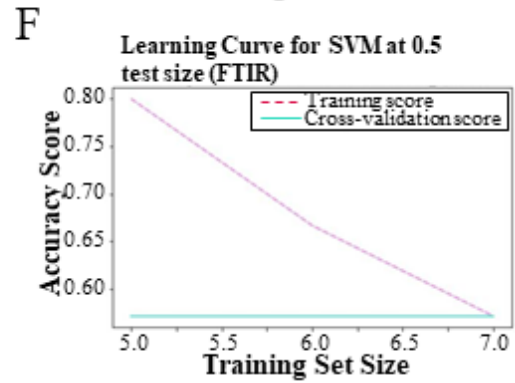

**FIGURE 4. ML PERFORMANCE ON BASELINE-CORRECTED FTIR SPECTRA WITH 5-FOLD CROSS-VALIDATION (CV). (A) ROC CURVE OF ADABOOSTCLASSIFIER ON BASELINE CORRECTED FTIR SPECTRA.**
(B) **CV CURVE FOR ADABOOST CLASSIFICATION ON BASELINE CORRECTED FTIR.** CV score: 100.00 $\pm$0.00%. (C)**ROC CURVE FOR DECISION TREES ON RAW FTIR SPECTRA**. (D) **CV CURVE FOR DECISION TREES ON RAW FTIR SPECTRA.** CV score of 90.00 ± 20.00%. (E) **ROC CURVE FOR SVM ON BASELINE CORRECTED FTIR SPECTRA.** 57.14 % classification accuracy and MSE of 0.4. The results confirm again that the baseline correction has poorer classification performance than the raw FTIR spectra. The f1 scores of the cancer and healthy groups' classification were found to be 0.73 and 0.00, respectively. **(F) CV CURVE OF SVM PERFORMANCE ON BASELINE CORRECTED FTIR SPECTRA.** CV accuracy: 75.00 ± 25.00%.

**TABLE 2.** ML CLASSIFICATION PERFORMANCE ON BASELINE-CORRECTED SPECTRA.

The performance metrics of the three ML classifiers are shown for the baseline-corrected Raman spectra (both reduced frequency space and complete spectra) and FT-IR spectra. The Classification accuracy, five-fold cross-validation score (CV), mean square error (MSE), and area under the curve (AUC) are summarized in the table. Feature ranking was not provided since performances are model-dependent and thus, hyperparameter-dependent (i.e., NP-hard problem).

| ML Algorithm | Statistical Measure | Raman (Selected Frequencies, Reduced) | Raman (Complete) | FT-IR |
|---|---|---|---|---|
| **AdaBoost RF** | Accuracy (%) | 83.33 | 88.89 | 80.00 |
| | CV-Score (%) | 83.33 ± 21.08 | 90.00 ± 20.00 | 100.00 $\pm$0.00 |
| | MSE | 0.167 | 0.111 | 0.2 |
| | AUC | 0.90 | 0.92 | 0.82 |
| **Decision Trees** | Accuracy (%) | 100.00 | 80.00 | 100.00 |
| | CV-Score (%) | 30.00 ± 24.45 | 60.00 ± 24.49 | 90.00 ± 20.00 |
| | MSE | 0.00 | 0.20 | 0.00 |
| | AUC | 1.00 | 0.81 | 1.00 |
| **SVM** | Accuracy (%) | 100.00 | 66.66 | 57.14 |
| | CV-Score (%) | 90.00 ± 20.00 % | 80.00 ± 24.49 | 75.00 ± 25.00 |
| | MSE | 0.00 | 0.333 | 0.4 |
| | AUC | 1.00 | 0.81 | 0.50 |

## 4. DISCUSSION

ML is emerging as a powerful tool in data science to help identify patterns in complex systems. For instance, ML algorithms applied on Intraoperative Raman spectroscopy can distinguish brain cancer cells from normal brain tissue (including both invasive and dense cancers) with an accuracy, sensitivity, and specificity > 90% (Jermyn et al., 2015). Background subtraction algorithms, feature extraction techniques and autofluorescence removal algorithms can be employed to distinguish the cancer cells' Raman bands from those of healthy tissues (Brusatori et al., 2017; Zhao et al., 2007). ML algorithms provide a robust computational platform to detect characteristic signatures and patterns in the complex spectra of cancer EVs, reshaping our understanding of these nanoscale communication networks (Uthamacumaran, 2020). Our study is a proof-of-concept demonstration of the promises machine intelligence holds in computational medicine and systems oncology.

Recent studies have demonstrated the longitudinal sampling of patient liquid biopsy-derived cell free DNA (cfDNA) and EVs/exosomes as potent biomarker discovery tools for early cancer detection and disease-prognostic screening. AI is paving minimally invasive, cancer screening with liquid-biopsy derived exosomes characterization. To illustrate, simple ML algorithms such as binomial classifiers (GLMnet) with the caret R package, can be trained to classify tumor-associated differential methylome patterns in cell-free plasma DNA (Shen et al., 2018). In a longitudinal study, Chen et al. (2020) demonstrated that PanSeer, a non-invasive blood test assessing circulating tumor DNA (ctDNA) methylation profiles, can detect five common types of cancers in 88% of cancer-diagnosed patients (n= 110) with a specificity of 96%. The study also revealed that aberrant cancer-specific ctDNA methylation patterns may be used as early cancer detection tools, by suggesting a > 90% detection in asymptomatic patients (n = 93) up to four years before cancer diagnosis, with 91% sensitivity. PanSeer utilizes an ensemble logistic regression classifier, coupled with covariate statistical analyses to perform these differential pattern recognitions of ctDNA methylome signatures in early cancer screening (Chen et al., 2020). More recently, the use of Deep Learning neural networks, a powerful subset of ML algorithms, are emerging as promising tools applicable in cancer diagnostics and screening. For instance, DISMIR, a deep learning algorithm, was shown capable of integrating the differentially methylated regions of ctDNA via whole-genome bisulphite sequencing (WGBS) as an ultra-sensitive and non-invasive cancer detection tool (Li et al., 2021). The central disadvantage within these AI-based cancer screening approaches using cell-free DNA (cfDNA) methylome signatures is that they are computationally and financially expensive, labor intensive, and may not be easily accessible to patients. A cheaper, quicker, and robust alternative is EV-based AI-driven biomarker discovery and pattern detection, as has been demonstrated in our pilot study.

Recently, EVs from stem cells were analyzed using Raman spectroscopy (Gualerzi et al., 2019). Following baseline correction and normalization of the obtained Raman spectra, cluster analysis and interpolation found that adult stem cell-derived EVs grouped differently when subjected to simple dimensionality reduction techniques like principal component analysis-linear discriminant analysis (PCA-LDA) (Gualerzi et al., 2019). Importantly, it was demonstrated that the PCA clustering of EVs from different stem cell types depended also on the purity of the EVs. Therefore, purification technique used to extract and isolate the EVs from the liquid biopsies, namely size-exclusion column chromatography or differential ultracentrifugation, plays a critical role in the cluster separation of EVs Raman spectra.

Similarly, surface-enhanced Raman scattering (SERS) signals of exosomes from normal and NSCLC (Non-small-cell lung carcinoma) cancer cells on gold-nanoparticle substrates were performed by Shin et al. (2018). The exosomes/EVs were isolated using size-exclusion column chromatography techniques. SERS provides an enhancement of the Raman scattering signals by use of metallic, nanostructured substrate. As such, even low concentrations of biomolecules can be detected using the SERS technique. The SERS Raman spectra of cancerous exosomes showed unique vibrational peaks distinguishing NSCLC exosomes from healthy clusters when subjected to spectral decomposition by Principal Component Analysis (PCA) (Park et al., 2017; Shin et al., 2018; Rojalin et al., 2019).

Shin et al. (2018) analyzed 25 spectra for each of the HPAEC (normal), H1299, and PC9 (NSCLC lung cancer cell lines) cell-derived exosomes, as well as the PBS control (pure buffer). PCA clustering well separated the cancer exosomes from the controls. Additionally, the cancerous exosomes tended to be located on the positive side of the PCA loading, which can be further screened for patient-specific biomarkers. Furthermore, correlation analysis on the Raman bands of exosome protein markers identified EGFR as the most unique Raman band for NSCLC exosomes (Shin et al., 2018). Similar findings with peak fitting algorithms and the multivariate curve resolution- alternating least squares (MCR-ALS) algorithm on Raman spectra were used to cluster-classify pancreatic cancer EVs on gold-nanoparticle plated SERS substrates (Banaei et al., 2017).

There exists many up-to-date systematic related works demonstrating the use of ML algorithms such as Deep Learning networks in diagnostic screening and patient-sera biomarker discovery (Zhou et al., 2021; Zhang et al., 2021). More recently, Deep Learning-based spectroscopic analysis of liquid-biopsy derived exosomes has shown > 90% sensitivity and accuracy in cancer detection (Shin et al., 2020). Exosomes are a subset of lipid-bound EVs found in the 30-100 nm size range circulating in patient blood sera. Shin et al. (2020) trained a residual neural network (Resnet)-based deep learning model, with the surface-enhanced Raman spectra (SERS) of cancer exosomes and healthy plasma cell line exosomes, which the algorithm classified with a 95% accuracy. When the Deep Learning algorithm was assessed on the SERS signals of 43 patients-derived exosomes, including stage I and II lung cancer patients, the algorithm predicted lung cancer with an AUC of 0.912 for the whole cohort and stage I patients with an AUC of 0.910. These findings strongly demonstrate that the pairing of liquid-biopsy derived cancer EVs with AI may pave early-stage cancer detection with high sensitivity and specificity (Shin et al., 2020).

Romano et al. (2020) used FTIR spectroscopy in the mid-Infrared (mid-IR) range to study exosomes released from human colorectal adenocarcinoma HT-29 cancer cells cultured in different media can be classified using PCA-LDA. Another recent study showed that neural networks optimized with ML algorithms such as principal component analysis– linear discriminant analysis (PCA-LDA) and binary classifiers like SVMs (Support Vector Machines) can distinguish oral cancer patients from healthy individuals by characteristic signatures in the Fourier transform infrared (FTIR) spectroscopy of their salivary exosomes (Zlotogorski-Hurvitz et al., 2019). Thus far, however, FTIR characterization of the complex spectra of patient-derived heterogenous exosomes (EVs) remain uninvestigated. Taken together, these findings collectively suggest simple ML algorithms coupled to spectroscopy techniques can pave the accurate detection and classification of cancers from patient-derived liquid biopsies. Our efforts and results as presented hereby may pave minimally invasive cancer screening with liquid-biopsy derived EVs characterization.

We also showed the emergence of crystals on the prepared EVs sample slides (See Figure 6 in Appendix). Caution must be taken to ensure the Raman laser is focused on the EVs solution and not these crystals whose origin is unknown. Raman imaging systems are thence useful in this procedure. The crystals may have emerged through the PBS buffer crystallization upon drying effect, and potentially catalyzed by the interaction with the incident laser beam (higher amounts of pyramids were present post-

## ML algorithms well-classified Baseline-corrected Raman spectra with a near 90% classification accuracy.

We demonstrated that the Raman spectra at a selected frequency region between 1800-1940 cm$^{-1}$ wavenumbers show several peaks distinguishing healthy from cancer EVs spectra. This selected frequency range corresponds to C=C double bonds and C=O double bonds (strong signals for Raman in this region) and were shown to enhance the classification accuracy of ML classifiers on the Raman spectra. However, their exact structural identity remains unelucidated. There are no major differences distinguishable at the level of the Raman peak assignment. This further supports that our traditional approach of peak assignments and molecular bond characterization are vastly insufficient for the study of complex systems such as cancer EVs. Although no visible peak differences are observed, the ML classifiers were able to distinguish the cancer EVs from healthy EVs via pattern recognition.

The spectra of all samples and their Raman/FTIR peak assignment to molecular bonds is provided in the Appendix. Additional information is also provided in the supplementary information. The alignment of the multiple spectral measurements for each sample indicates there are no differences in the spectra due to sample preparation, impurities, or differences due to optical effects present (Figures S1-S6 in Supplementary Information). We ruled out differences seen due to spatial heterogeneity instead of differences between samples with repeated measurements. The FTIR data of both cancer samples 426 (colorectal cancer) and 549 (hepatocellular cancer) revealed lipid bands in the 2850-3010 cm$^{-1}$ region (corresponding to C-H stretches) (Figure 1A- D). The 3200-3400 cm$^{-1}$ stretch corresponds to the OH stretch. Some peaks which are tentatively present only within the cancer FTIR spectra were present at 1120 cm$^{-1}$, and at 3070 cm$^{-1}$ (Figure 1E and 3F).

When the Raman spectra was reduced to only the 1800-1940 cm$^{-1}$ wave number region, a better PCA separation was observed in 18 samples with 27 intensity data points in each (one control was removed as an outlier due to low intensity counts) (Figure 2). The cancer spectra, although diverse and heterogeneous (there are four different cancer subtypes present) seem to cluster in the left quadrant of the PCA space. There is still one green data point (healthy spectrum) clustered with the cancer spectra. In the following ML classification results, *reduced (or selected) frequency* denotes Raman spectra being processed to only the 1800-1940 cm$^{-1}$ region of interest (unless otherwise specified, as in the case for the Decision Trees). With further detailed spectral analysis, additional spectral ranges can be included in downstream analysis. A correct artifact filtering and peak fitting of the Raman spectra is required to primarily focus on the relevant spectral information only and achieve a more efficient PCA classification. Even simple dimensionality reduction techniques like PCA provide a quick tool to characterize and distinguish serum derived EVs from cancer patients with those of healthy individuals.

Figure 3 displays the ML results on the baseline corrected Raman spectra. A classification accuracy of 88.89% and MSE of 0.111 was observed for the AdaBoost RF classifier on baseline corrected Raman spectra. The f1 scores of the cancer and healthy group classification were 0.91 (i.e., precision of 1.00 and recall of 0.83) and 0.86 (i.e., precision of 0.75 and recall of 1.00), respectively corresponding to an AUC of 0.92 (Figure 6A). A classification accuracy of 83.33% and MSE of 0.167 was found for the AdaBoost classifier on baseline corrected Raman spectra within the reduced frequency range. The f1 scores of the cancer and healthy group classification were 0.89 (i.e., precision of 1.00 and recall of 0.80) and 0.67 (i.e., precision of 0.50 and recall of 1.00), respectively corresponding to an AUC of 0.90 (Figure 6C). As shown, in comparison to Figure 7 in Appendix (i.e., no baseline correction), the ML performance enhanced by baseline correction of the spectra. Hence, some of the essential features of the raw Raman spectra can be removed by the baseline correction or perhaps the quality of the baseline correction. Although, the classification accuracy remains roughly the same for both baseline-corrected spectra in the complete range and reduced frequency range, their precision has changed especially for the healthy controls (Figure 3D). Similar classification accuracies were obtained for the Decision Trees and SVM algorithms as well (Figure 3E-H). Their classification accuracy was enhanced to 100.00% on the reduced frequency spectra (1800-1940 $cm^{-1}$). However, the improved classification accuracy was accompanied by a lowered cross-validation score (i.e., increased standard deviation). The ML performance scores on the baseline corrected spectra are summarized in table 2. These results collectively confirm that the baseline correction improves the ML classification accuracy in detecting patterns distinguishing cancer spectra from healthy spectra.

***Preliminary findings show ML algorithms better predict cancer EVs from Raman spectra than from FTIRspectra.***

As demonstrated in Figure 4, the FTIR was used as a complementary verification of the Raman data. AdaBoost Classifier was tested on N = 14 baseline corrected FTIR spectra, 857 points each spanning from 698 to 4000 $cm^{-1}$. An 80.00% classification accuracy and MSE of 0.2 was observed. The f1 scores of the cancer and healthy groups classification were 0.80 for both leading to an AUC of 0.83. Our findings show the raw data (no baseline correction) performed better than the baseline corrected FTIR spectra. The opposite trend was observed for the Raman spectra where a better classification accuracy was seen overall with the baseline corrected spectra (Figure 4A). The classification accuracy using the Decision Trees algorithm for a randomly selected frequency (at 698.22975 $cm^{-1}$) was found to be 100.00 % with a MSE of 0.0. The specificity and sensitivity were 1.00. Our findings repeatedly confirm Decision Trees at selected frequencies may be robust predictors of distinguishing healthy from cancer spectra in both cases, the Raman, and FTIR methods (Figure 4C). The comparison of various statistical measures to assess the performance of the three ML classifiers are provided in table 2.

Our results demonstrate vibrational spectroscopies coupled with basic machine learning algorithms may be a robust tool for the detection of cancer in patient-derived liquid biopsies by the characterization of EVs. One limitation of our study is the sample size. Hence, the inclusion of a larger sample size of patients with cancers of different origins and grades/stages is needed to enhance its impact in clinical medicine. Regardless, our findings serve as a pilot study in the use of machine learning/ artificial intelligence in liquid-biopsy based early cancer detection and prognostic screening. Another limitation is the presence of lipoproteins in our patient- derived EVs samples. Lipoproteins will co-isolate with the EVs, even after flotation. Size-exclusion chromatography following the ultracentrifugation could have partially resolved this issue. However, given the lipoproteins were present in all patients, and our algorithms distinguished between cancer and healthy patients with > 90% accuracy, patterns are consistently observed between the two patient groups which could have not been affected by the lipoproteins.

**Prospective studies:** Our pilot study demonstrates that the longitudinal sampling of sera-derived patient EVs coupled with artificial intelligence could provide clinically-relevant cancer screening tools in the emerging field of computational precision oncology. Although the AdaBoost Random Forest classifier accurately distinguished between healthy and cancer EVs, the performances of ML algorithms highly rely on the model hyperparameters and thresholds. Our findings can be further improved using SERS (Surface Enhanced Raman Spectroscopy) on patient samples with a uniform gold substrate to obtain a better resolution of Raman vibrational peaks. These preliminary findings with N= 18 analysis strongly suggest that Raman spectroscopy in combination with machine learning classifiers can be used to develop sensitive liquid biopsies for early cancer detection in screened patients.

PCA was sufficient as a feature selection and data representation method to accurately cluster-classify cancer spectra from healthy spectra. It can be used as an additional layer of information processing in the training of more complex ML algorithms such as Deep Learning Networks (Shin et al., 2020) in prospective studies. Nonlinear feature selection was not performed since the linear dimensionality reduction by PCA was sufficient to well-distinguish the two patient groups as shown in Figure 1. However, nonlinear feature extraction measures such as nonlinear dimensionality reduction methods/manifold learning to enhance the data representation space (e.g., nonlinear neighborhood component analysis (NNCA), k-means clustering, Uniform Manifold Approximation and Projection (UMAP), Gaussian Mixture Models (GMM), Gaussian Process latent variable models (GPLVM), diffusion maps, Isomap, multidimensional scaling (MDS), etc.), multifractal analysis, spectral clustering algorithms, and spectral power analyses (i.e., Fast-Fourier Transform) could be used in future studies on the EV spectra to identify discriminants of cancer and healthy spectra. PCA clustering and peak fit analyses can be added as additional features for pattern recognition training in multi-layered neural network architectures such as Deep Learning classifiers.

Additionally, we could use multivariate information-theoretics on the EV spectra such as partial information decomposition, mutual information, or the algorithmic complexity (K-complexity) of identified characteristic spectral peaks distinguishing cancer EVs from healthy EVs (Chan et al., 2017; Zenil et al., 2019). Graph-spectral complexity analysis can assess the K-complexity of the graph-spectra using the 2D-Block Decomposition Method (BDM) algorithm (Soler-Toscano et al., 2014; Zenil et al., 2014). Further, while most of our current approaches rely on statistical machine learning methods to identify correlations in complex datasets as disease indicators or biosignatures, causal inference and causal pattern discovery remain primitively developed in biomedical sciences (Uthamacumaran, 2020). AI and algorithms provide a host of causal inference tools in the time-series classification of complex biological spectra/datasets which should be exploited in prospective EVs characterization studies with time-series spectroscopic techniques (Zenil et al., 2019; Yan et al., 2020). Recurrent Neural Networks such as Reservoir Computing and liquid neural networks are available as tools for causal pattern discovery in complex time-series datasets which could be exploited in time-series EVs spectroscopic measurements and longitudinal blood monitoring (Maass et al., 2002; Verstraeren et al., 2007; Pathak et al., 2018; Hasani et al., 2020). The emerging paradigm of quantum machine learning may also provide a novel set of tools for biomarker discovery and pattern detection.

Moreover, we should exploit other types of spectroscopic techniques and light-matter interaction interfaces including mass spectrometry/CyTOF (for proteomic profiling of EVs) (Hoshino et al., 2020), fluorescence spectroscopies, and time-series spectroscopies in liquid biopsy/EVs characterization by AI. As discussed, tumor specific cfDNA methylome signatures can be used to detect cancer tissue of origin (molecular subtyping) and early cancer screening. The differential methylation patterns and somatic mutations of cfDNA can be used to train machine learning classification algorithms like Deep Learning networks for cancer diagnosis. However, these high-throughput sequencing methods are both financially and computationally expensive and present a fundamental barrier in accessibility for patient-centered precision medicine. Our Raman spectroscopy-based AI approach is a quick, robust, non-invasive, and efficient algorithm with a low computation time and relatively cheap screening method. The disadvantage of our study remains the low patient size, and future studies should train the ML classifiers with an increased patient size with different stages and molecular subtypes of cancers. However, as demonstrated in Table 2, the statistical performance of our classifiers are relatively high due to classifier regularization (to prevent overfitting) and hyperparameter optimization of the models (e.g., the boosting method for RF to optimize bias-variance trade-off).

The scope of artificial/applied intelligence in improving precision medicine via biomarker discovery was demonstrated by our application of AI in patient liquid-biopsy derived EV characterization in cancer screening. Despite the small sample size, our study is to be treated as a pilot study in demonstrating for the first time the use of the presented algorithms in accurately distinguishing cancer and healthy patient derived EVs, as further supported by their high statistical performance measures (Table 2). Large-scale studies with Deep Learning Network classifiers on different stages of cancers and an increased patient cohort are suggested as future studies of our applied intelligence model.

Lastly, it should be further noted that our presented AI-driven liquid-biopsy system has potential impact in other related application domains. Some pertinent examples in medical diagnosis include fuzzy systems applications such as the use of cosine similarity measures demonstrated in prostate cancer screening (Fan et al., 2018; Cui et al., 2020). As such, we propose future classification tasks should exploit fuzzy systems and soft computing measures in cancer screening. Multi-features-based ensemble learning methods and similar expert AI systems also show great promise in cancer screening as demonstrated by our AdaBoost Random Forest classifier. For example, ensemble methods based on symbolic aggregate approximation such as TBOPE can provide an effective mean feature pattern-recognition and time-series trend extraction method for prospective time-series based spectroscopies in EVs characterization (Bai et al., 2021).

Knowledge extraction and knowledge representation are usually domain-specific. However, these suggested AI systems can extract and represent knowledge from multiplicity of domains and hence show generalized application to other complex problems in precision medicine (Cui et al., 2020; Bai et al., 2021). We also suggest that statistical data-driven liquid-biopsy screening approaches, such as the AI system presented in our pilot study, should be combined with physics model-driven and causal discovery methods (which require time-series spectroscopies). The temporal behaviors and causal patterns of EVs dynamics/vibrations should be quantified in the foreseeable future along with the spatial profiles investigated herein. We should extend our studies not only to higher-orders of statistically-driven AI approaches such as deep learning space but other types of domain AI systems such the above-mentioned liquid cybernetics for causal inference (e.g., liquid neural networks, reservoir computing, dynamical systems approaches, etc.) and neuro-symbolic AI to make significant progress in forecasting/predicting disease dynamics in computational systems oncology (Maass et al., 2002; Hasani et al., 2020). Again, the scope of quantum machine learning, which remains at its infancy, should also be explored in prospective domain applications of our applied intelligence system.

## 5. CONCLUSION

Various classification algorithms from the paradigm of statistical machine learning were used here as applied intelligence to solve a real-life complex problem: biomarker discovery for early-cancer detection. Our approach is both a computationally and financially cheap, robust, quick, and accurate diagnostic tool in early cancer detection and prognostic screening. The presented findings strongly demonstrate that the ML classifiers can accurately distinguish cancer EVs from those of healthy patients using both vibrational spectroscopy techniques. Traditional peak assignment approaches are insufficient in assessing the complex spectra of cancer EVs and distinguishing them from those of healthy controls. Our study demonstrates basic ML classifiers can find patterns which traditional approaches fail to reconcile. It must be reminded that these are heterogeneous complex adaptive systems since the EVs were acquired from the blood sera of patients. A wide repertoire of EVs from different tissues within the patient are expected in each sample (i.e., deriving from inflammatory and other organismal cells). Previous studies only investigated cancer EVs obtained from cancer cell lines which reduce the degree of complexity in comparison to our samples. However, regardless, the ML algorithms identified Raman signatures distinguishing the two groups with high degrees of statistical accuracy, sensitivity, and specificity. The ML performance on Raman spectra had a better classification accuracy than those on FTIR spectra. Our study displays the impact and scope of interdisciplinary ML applications of EVs in improving precision healthcare and computational systems medicine.

**DISCLOSURE STATEMENT:** The authors report no conflict of interest.

**FUNDING STATEMENT:** Giuseppe Monticciolo financially supported the research and the experiments described in this paper. The funder had no role in study design, data collection and analysis, decision to publish, or preparation of the manuscript.

**DATA AVAILABILITY STATEMENT:** All data generated and analyzed during this study are included in this manuscript and in its Appendix files. Sample Raman spectra and Google Colab codes for the ML classifiers are available in our Github link: https://github.com/Abicumaran/Exosomes-ML-Classifiers

**ETHICS APPROVAL STATEMENT:** Ethics approval and consent to participate Patients recruited for this study underwent an informed and written consent for blood collection in accordance to a protocol approved by the Ethics Committee of the McGill University Health Centre (Reference. MP-37-2018- 3916 and 10–057-SDR).

**PATIENT CONSENT STATEMENT:** The authors declare patient consent was granted for the study.

**AUTHOR CONTRIBUTIONS:**
AU performed the machine learning algorithms, co-wrote, and edited the manuscript.
SE carried out the spectroscopy measurements.
MA extracted and purified the patient-EVs, co-wrote, and edited the manuscript.
MBR performed the baseline corrections and spectral peak fit analysis.
ZHG co-supervised the project.
GA co-supervised the project, co-wrote, and edited the manuscript.

**REFERENCES**

1) Cancer. World Health Organization. https://www.who.int/news-room/fact-sheets/detail/cancer (visited on June 2021)

2) Théry, C. et al. Minimal information for studies of extracellular vesicles 2018 (MISEV2018): a position statement of the International Society for Extracellular Vesicles and update of the MISEV2014 guidelines. J. Extracell. Vesicles. 7: 1535750 (2018)

3) Uthamacumaran, A. A Review of Complex Systems Approaches to Cancer Networks. Complex Systems 29(4): 779-835 (2020)

4) Samuel, P., Fabbri, M., & Carter, D. R. F. (2017). *Mechanisms of Drug Resistance in Cancer: The Role of Extracellular Vesicles. PROTEOMICS, 17(23-24), 1600375.* doi:10.1002/pmic.201600375

5) Ramakrishnan, V., Xu, B., Akers, J., Nguyen, T., Ma, J., Dhawan, S., … Chen, C. C. (2020). *Radiation-induced extracellular vesicle (EV) release of miR-603 promotes IGF1-mediated stem cell state in glioblastomas. EBioMedicine, 55, 102736.* doi:10.1016/j.ebiom.2020.102736

6) Fontana, F., Carollo, E., Melling, G. E., & Carter, D. (2021). Extracellular Vesicles: Emerging Modulators of Cancer Drug Resistance. *Cancers*, *13*(4), 749. https://doi.org/10.3390/cancers13040749

7) Guo, Y. et al., Effects of Exosomes on Pre-metastatic Niche Formation in Tumors. MolecularCancer, 18:39 (2019)

8) Abdouh M, Zhou S, Arena V, Arena M, Lazaris A, Onerheim R, Metrakos P, Arena GO. Transfer of malignant trait to immortalized human cells following exposure to human cancer serum. J Exp Clin Cancer Res. 33:86. 2014

9) Abdouh M, Hamam D, Arena V, Arena M, Alamri H, Arena GO. Novel blood test to predict neoplastic activity in healthy patients and metastatic recurrence after primary tumor resection. J Circ Biomark. 5. doi: 10.1177. (2016).

10) Abdouh M, Hamam D, Gao ZH, Arena V, Arena M, Arena GO. Exosomes isolated from cancer patients' sera transfer malignant traits and confer the same phenotype of primary tumors to oncosuppressor-mutated cells. J Exp Clin Cancer Res. 36(1):113. 2017.

11) Abdouh M, Floris M, Gao ZH, Arena V, Arena M, Arena GO. Colorectal cancer-derived extracellular vesicles induce transformation of fibroblasts into colon carcinoma cells. J Exp Clin Cancer Res. 38(1):257. 2019. / Abdouh M, Gao ZH, Arena V, Arena M, Burnier MN, Arena GO. Oncosuppressor-Mutated Cells as a Liquid Biopsy Test for Cancer-Screening. Sci Rep. 9(1):2384 (2019).

12) Abdouh M, Tsering T, Burnier JV, de Alba Graue PG, Arena G, Burnier MN. Horizontal transfer of malignant traits via blood-derived extracellular vesicles of uveal melanoma patients. Invest. Ophthalmol. Vis. Sci. 61(7):2835 (2020)

13) Arena GO, Arena V, Arena M, Abdouh M. Transfer of malignant traits as opposed to migrationof cells: A novel concept to explain metastatic disease. Med Hypotheses. 100:82-86 (2017)

14) Hamam D, Abdouh M, Gao ZH, Arena V, Arena M, Arena GO. Transfer of malignant trait to BRCA1 deficient human fibroblasts following exposure to serum of cancer patients. J Exp ClinCancer Res. 35:80. 2016.

15) Steinbichler, T.B. et al., Therapy resistance mediated by exosomes. Molecular Cancer 18:58(2019)

16) Keklikoglou, I. et al., Chemotherapy elicits pro-metastatic extracellular vesicles in breast cancer models. Nature Cell Biology 21(2): 190-202 (2019)

17) Camussi, G. et al., Exosome/Microvesicle-Mediated Epigenetic Reprogramming ofCells. American Journal of Cancer Research 1(1): 98–110 (2011)

18) Zhou, S. et al., Reprogramming Malignant Cancer Cells toward a Benign Phenotype following Exposure to Human Embryonic Stem Cell Microenvironment. PloS One 12,1 e0169899 (2017)

19) Zhao Z, Fan J, Hsu Y-M, Lyon CJ, Ning B, Hu TY. Extracellular vesicles as cancer liquid biopsies:from discovery, validation, to clinical application. Lab Chip. 2019;19(7):1114-1140. doi: 10.1039/c8lc01123k.).

20) Ember, K. et al., Raman spectroscopy and regenerative medicine: a review. Npj Regenerative Medicine, 2(1):12 pp.1-12. (2017)

21) Smith, E. and Dent, G. Modern Raman Spectroscopy - A Practical Approach (John Wiley andSons, Ltd, England, 2005)

22) Larkin, P. Infrared and Raman Spectroscopy: Principles and Spectral Interpretation (Elsevier,2011)

23) Brusatori, M. et al. Intraoperative Raman Spectroscopy. Neurosurgery clinics of NorthAmerica 28(4): 633–652 (2017)

24) Pedregosa, F. et al., Scikit-learn: Machine Learning in Python, JMLR 12: 2825-2830 (2011)

25) Breiman, L. Random Forests. Machine Learning, 45(1): 5–32 (2001)

26) Bishop, C.M. Pattern Recognition and Machine Learning (Springer, 2006)

27) Jermyn, M. et al., Intraoperative Brain Cancer Detection with Raman Spectroscopy in Humans. Science Translational Medicine, 7(274): 274ra19 (2015)

28) Zhao, J. et al., Automated Autofluorescence Background Subtraction Algorithm for Biomedical Raman Spectroscopy. Applied Spectroscopy 61(11): 1225-1232 (2007)

29) S.Y. Shen, R. Singhania, G. Fehringer, A. Chakravarthy, M. Roehrl, D. Chadwick, P.C. Zuzarte et al., "Sensitive tumor detection and classification using plasma cell-free DNA methylomes,"Nature **563** (7732), 2018 pp. 579-583. https://doi.org/10.1038/s41586-018-0703-0.

30) Chen, X., Gole, J., Gore, A. et al. Non-invasive early detection of cancer four years before conventional diagnosis using a blood test. Nat Commun **11,** 3475 (2020). https://doi.org/10.1038/s41467-020-17316-z

31) Li, J., Wei, L., Zhang, X., Zhang, W., Wang, H., Zhong, B., Xie, Z., Lv, H., & Wang, X. (2021). DISMIR: Deep learning-based non-invasive cancer detection by integrating DNA sequence and methylation information of individual cell-free DNA reads. *Briefings in bioinformatics*, bbab250. Advance online publication. https://doi.org/10.1093/bib/bbab250.

32) Gualerzi, A. et al., Raman spectroscopy as a quick tool to assess purity of extracellular vesicle preparations and predict their functionality. Journal of extracellular vesicles, 8(1):1568780 (2019)

33) Shin, H. et al., Correlation between Cancerous Exosomes and Protein Markers Based on Surface-Enhanced Raman Spectroscopy (SERS) and Principal Component Analysis (PCA). ACS Sensors 3(12):

2637–2643 (2018)

34) Park, J. et al., Exosome classification by Pattern analysis of surface-enhanced Ramanspectroscopy data for lung cancer. Analytical Chemistry 89(12): 6695–6701 (2017)

35) Rojalin, T. et al., Nanoplasmonic Approaches for Sensitive Detection and MolecularCharacterization of Extracellular Vesicles. Frontiers in Chemistry 7:279 (2019)

36) Banaei, N. et al., Multiplex detection of pancreatic cancer biomarkers using a SERS-based immunoassay. Nanotechnology 28(45): 455101 (2017)

37) Zhou, J., Zhang, X., and Jiang, Z. "Recognition of Imbalanced Epileptic EEG Signals by a Graph-Based Extreme Learning Machine," Wireless Communications and Mobile Computing, vol. 2021 (2021).

38) Zhang, J., Yu, Y., Fu, S., and Tian, X. "Adoption value of deep learning and serological indicators in the screening of atrophic gastritis based on artificial intelligence," The Journal of Supercomputing, pp. 1-20 (2021).

39) Shin, H., Oh, S., Hong, S., Kang, M., Kang, D., Ji, Y. G., Choi, B. H., Kang, K. W., Jeong, H., Park, Y., Hong, S., Kim, H. K., & Choi, Y. (2020). Early-Stage Lung Cancer Diagnosis by Deep Learning-Based Spectroscopic Analysis of Circulating Exosomes. *ACS nano*, *14*(5), 5435–5444. https://doi.org/10.1021/acsnano.9b09119

40) Romano, Sabrina; Di Giacinto, Flavio; Primiano, Aniello; Mazzini, Alberto; Panzetta, Claudia; Papi, Massimiliano; Di Gaspare, Alessandra; Ortolani, Michele; Gervasoni, Jacopo; De Spirito, Marco; Nocca, Giuseppina; Ciasca, Gabriele (2020). Fourier Transform Infrared Spectroscopy asa useful tool for the automated classification of cancer cell-derived exosomes obtained under different culture conditions. Analytica Chimica Acta, 1140:219-227. doi:10.1016/j. aca.2020.09.037

41) Zlotogorski-Hurvitz, A. et al. FTIR-based spectrum of salivary exosomes coupled with computational-aided discriminating analysis in the diagnosis of oral cancer. Journal of Cancer Research and Clinical Oncology 145(3): 685–694 (2019)

42) Chan, T. E et al. Gene Regulatory Network Inference from Single-Cell Data Using Multivariate Information Measures. *Cell systems* vol. 5(3): 251-267.e3. (2017). doi:10.1016/j.cels.2017.08.014

43) Zenil, H., Kiani, N.A., Marabita, F, Deng, Y., Elias, S., Schmidt, A., Ball, G., and Tegnér, J. [2019] An Algorithmic Information Calculus for Causal Discovery and Reprogramming Systems, iScience 19: 1160-1172.

44) Soler-Toscano F., Zenil H., Delahaye J.-P. and Gauvrit N. (2014) Calculating Kolmogorov Complexity from the Output Frequency Distributions of Small Turing Machines. PLoS ONE 9(5): e96223.

45) Zenil H., Soler-Toscano F., Dingle K. and Louis A. (2014) Correlation of Automorphism Group Size and Topological Properties with Program-size Complexity Evaluations of Graphs and Complex Networks, Physica A: Statistical Mechanics and its Applications, vol. 404, pp. 341–358.

46) Yan, W., Li, G., Wu, Z., Wang, S., and Yu, P.S., "Extracting diverse-shapelets for early classification on time series," World Wide Web, vol. 23, pp. 3055-3081 (2020).

47) Maass, W. et al. [2002] Real-time computing without stable states: A new framework for neural computation based on perturbations. Neural Computation 14, 2531-2560.

48) Verstraeten, D. et al. [2007] An experimental unification of reservoir computing methods. Neural Networks, 20:391-403

49) Pathak, J., Hunt, B., Girvan, M., Lu, Z., & Ott, E. [2018]. Model-Free Prediction of Large Spatiotemporally Chaotic Systems from Data: A Reservoir Computing Approach. Phys. Rev. Lett. 120[2]: 024102.

50) Hasani, R. et al., Liquid Time-constant Networks. arXiv:2006.04439 [cs.LG] (2020)

51) Hoshino, A. et al., Extracellular Vesicle and Particle Biomarkers Define Multiple Human Cancers. Cell 182(4): 1044–1061.e18 (2020)

52) Géron, A., Hands-On Machine Learning with Scikit-Learn, Keras, and TensorFlow (Second Ed., O'Reilly Media, Inc, 2019)

53) Fan C, Fan E, Ye J. The Cosine Measure of Single-Valued Neutrosophic Multisets for Multiple Attribute Decision-Making. *Symmetry*. 10(5):154 (2018) https://doi.org/10.3390/sym10050154

54) Cui, WH., Ye, J. & Fu, J. Cotangent similarity measure of single-valued neutrosophic interval sets with confidence level for risk-grade evaluation of prostate cancer. *Soft Comput* **24,** 18521–18530 (2020). https://doi.org/10.1007/s00500-020-05089-y

55) Bai, B., Li, G., Wang, S., Wu, Z., & Yan, W. Time series classification based on multi-feature dictionary representation and ensemble learning. Expert Systems with Applications, 169: 114162 (2021). doi:10.1016/j.eswa.2020.114162

**APPENDIX**

1. **EVs isolated from serum displayed exosomes and microvesicles characteristics.**

   Based on the minimal information statement for the study of EVs set by the ISEV (International Society for Extracellular Vesicles) (Théry et al., 2018), we characterized the isolated EVs both physically and phenotypically. By using Western blot analysis, we observed that these vesicles expressed selective markers of EVs (i.e., Alix and TSG101) (Figure 5A). The highest expression levels of these markers were observed in fractions 3-5 (at iodixanol density of 1.107-1.13 g/ml). These fractions were subsequently pooled for further analyses. When assessed by NTA, isolated EVs displayed a mean diameter of 109 nm (range 59-145 nm) (Figure 5B). In addition, TEM analyses showed that the isolated EVs were round-shaped vesicles with a mean diameter of 90 nm (Figure 5C).

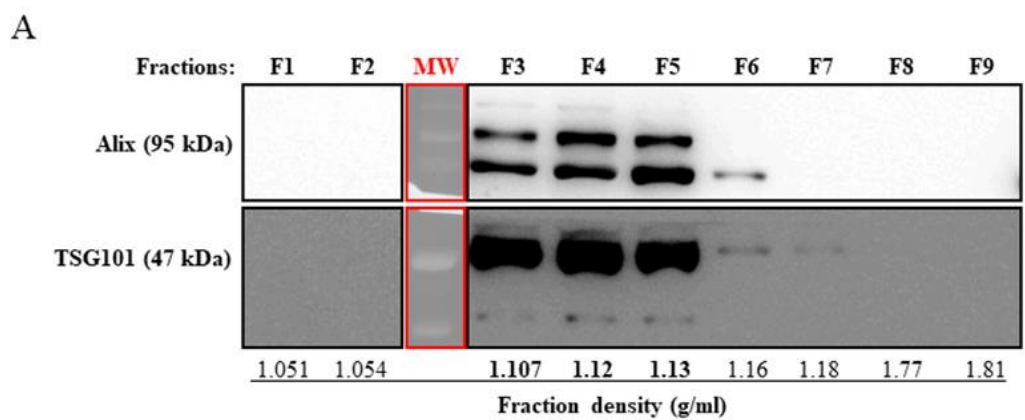
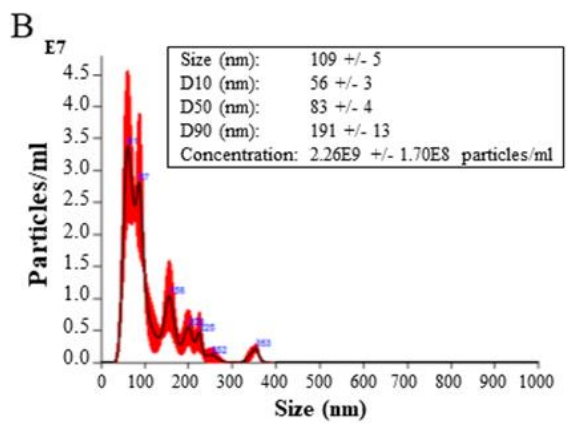
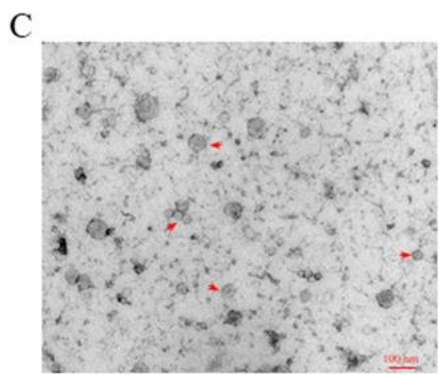

**FIGURE 5. ISOLATION AND CHARACTERIZATION OF PATIENT-DERIVED EVs.** (A) EVs were isolated as described under Materials and Methods using iodixanol (OptiPrep) gradient and ultracentrifugation. Proteins isolated from the different fractions were analyzed by Western blot for the expression of specific EVs markers. Note that the highest expression levels of EVs markers are located in fractions 3 to 5. These fractions were pooled for subsequent analyses. MW = Protein molecular weight marker. (B) NTA analysis of pooled fractions 3-5. (C) TEM micrograph of purified EVs (red arrowheads). Scale bar 100 nm.

2. **Renishaw Raman Spectroscope identifies rich spots of EVs on air-dried CaF$_2$ slides.** Spectra acquisition consisted of scanning at multiple pink spots on the prepared sample slides (Figure 6A). The spontaneous formation of pyramidal crystals on the slides are shown (Figure 6B).

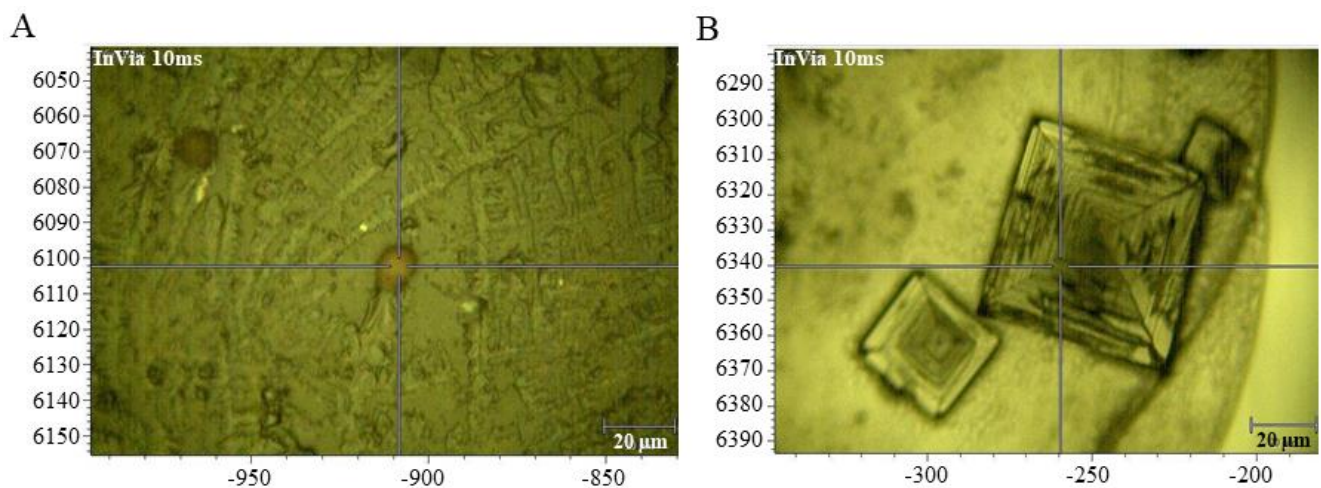

**FIGURE 6. EVs DETECTION BY SPECTROSCOPY.** (A) EVs SPOT ON RAMAN CONFOCAL MICROSCOPE (50X Objective). The axes represent the microscope field of view sizes in X and Y directions corresponding to the 50x lens that was used for measurement. The inset bar gives the field of view scale corresponding to the 50x objective lens. The image shows the EVs spot focused on the air-dried CaF$_2$ slide through the confocal microscope of the inVia Renishaw Raman system, with the scale bar of 20 μm as indicated by the scale bar on the bottom. These pink spots are regions enriched with EVs and provide a method to infer where the optimal Raman spectra are acquired. The yellow patchy regions correspond to the slide with higher concentrations of the PBS buffer. (B). CRYSTAL FORMATION. Pyramidal crystals spontaneously self-organized in the EVs rich regions (pink spots).

### 3. Machine Learning algorithms exhibit poorer performance in the classification of EVs on Raman spectra without baseline correction.

Various binary classification algorithms were trained on raw Raman spectra without baseline correction to assess their performance accuracy in distinguishing cancer samples from healthy controls (Figures 5A-J). All results collectively confirm that without baseline correction, ML algorithms exhibit poor predictive performance.

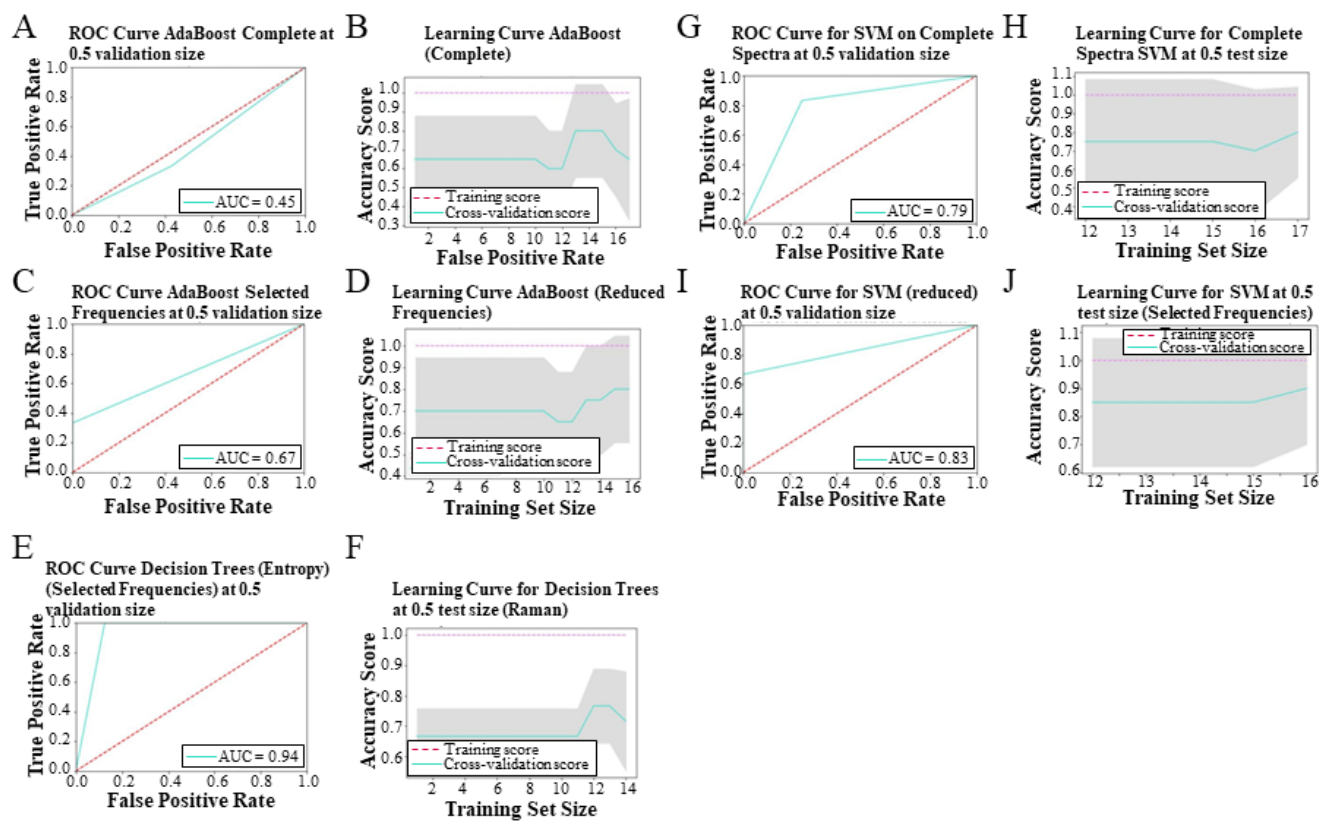

**FIGURE 7. ML PERFORMANCE ON RAW RAMAN SPECTRA WITHOUT BASELINE CORRECTION. (A). ADABOOST RANDOM FOREST (RF) ONWHOLE RANGE RAW RAMAN SPECTRA. (B) ADABOOST 5-FOLD CROSS-VALIDATION CURVE ON RAW RAMAN SPECTRA FOR WHOLE-RANGE. (C) ADABOOST RF CLASSIFICATION ON RAW RAMAN SPECTRA WITH REDUCED FRREQUENCY. (D) CV CURVE FOR ADABOOST ON REDUCED RAW RAMAN SPECTRA.** CV accuracy score: 20.00 ± 24.49%. **(E)ROC CURVE FOR DECISION TREES ON RAW RAMAN SPECTRA NO BASELINE. (F) CV CURVE FOR DECISION TREES PERFORMANCE ON RAW RAMAN SPECTRA (at 1808.25 cm$^{-1}$).** The CV accuracy score was determined as 60.00 $\pm$ 20.00 %. With an increased sample size of patients and hence, increased training datasets the performance of these algorithms can be better optimized for the intended impact of the presented findings. **(G) ROC CURVE FOR SVM CLASSIFICATION ON RAW RAMAN SPECTRA (COMPLETE). (H) CLASSIFICATION REPORT FOR SVM ON RAW RAMAN SPECTRA (FULL RANGE).** CV accuracy score: 80.00 ± 24.49 %**. (I) ROC CURVE FOR SVM PERFORMANCE ON RAW RAMAN SPECTRA WITH REDUCED FREQUENCY RANGE. (J) CV CURVE FOR SVM PREDICTIONS ON REDUCED SPECTRA.** CV accuracy score: 90.00 ± 20.00 %.

The AdaBoost RF classifier is a meta-estimator and an iterative ensemble learner available on Sci- kit-learn (*Python machine learning library*). The AdaBoost RF classifier was assessed on 1021 data points from 19 raw Raman spectra without baseline correction from the two classes (Figure 7). To assess the sensitivity/specificity of the ML predictions, receiver operating characteristic (ROC) curves are generated to show the diagnostic ability of the binary classifier with varied discrimination thresholds. As shown in Figure 5, the ROC curve for the AdaBoost classifier's performance on the raw Raman spectra (whole range) with a testing size of 0.5 (i.e., the ML is trained on 50% of the data and tested on 50%) is shown (Figure 7A). A 0.5 test size ensures stringent training conditions for the classification assessment. A smaller test size of 0.2 and 0.3 always showed greater classification accuracy. 50 tree estimators and a learning rate of 1.0 were kept as the default hyperparameters. Theclassification accuracy was 77.78% with a mean-square error (MSE) of 0.222. As seen, the area under the curve (AUC) was 0.45 indicating a poor classification accuracy. The turquoise curve shows the relationship between the true positive rate (TPR) and false positive rate (FPR). The closer the turquoise curve comes closer to the red dashed curve at 45 degrees of the graph plane, the less accurate the classifier's predictions, and lower AUC of the ROC curve. The ROC curve visually informs us the trade-off between the sensitivity (TPR) and the specificity (1-FPR) (Bishop, 2006). The f1 scores were 0.62 (i.e., a precision of 0.67 and recall of 0.57) for the cancer group and 0.29 (i.e., a precision of 0.25 and recall of 0.33) for the healthy groups. The F1 score of 1.00 indicates a perfect recall and precision. The F1-score is often used as a measure of statistical accuracy for binary classifiers in ML (Breiman, 2001; Geron, 2019).

The poor performance of the RF classifier in Figure 7 indicates the data must be filtered to a narrower spectral range or alternately, undergo a baseline correction, as indicated by the PCA plot of the reduced frequency space above. In binary classification, recall of the positive class is defined as sensitivity while the recall of the negative class is specificity. The precision is defined as the ratio tp/ (tp + fp) where tp is the number of true positives and fp the number of false positives. The recall is the ratio tp/(tp+fn) where fn is the number of false negatives. The recall denotes the ability of the classifier to find all the positive samples. The f-1 score defines the weighted harmonic mean of the precision and recall, where an f-1 score reaches its best value at 1 and worst score at 0. Here,the f1 score was found to be of 1.00 (i.e., precision and recall were 1.00) implying both, a 100% sensitivity and specificity.

Figure 7B displays the cross-validation learning curve corresponding to graph 7A. It shows that the AdaBoost classifier is not optimally tuned to predict the classes of the newly presented test datasets as indicated by the vast grey shaded region (indicates training uncertainty). The grey fill space on the plotdenotes the standard deviation of the training performance by the classifier as the training size increases. The broad range of grey fill indicates a heavier computational training is required for the classification accuracy to be optimized. The cross-validation score, or also known as out-of-sample testing, indicates the likelihood of the RF classifier's performance when new results are presented withthe current amount of training it has undergone. It is a validation technique to generalize the performance of the RF classifier to an independent dataset. The five-fold CV accuracy score was found to be 70.00 ± 24.49 %. The curve in turquoise corresponds to the CV score curve and the optimal training score of 1.00 is indicated by the dashed violet curve.

The RF classifier was assessed on the frequency reduced spectra (1800-1940 cm$^{-1}$ wavenumber region) of 18 samples (12 cancers and 6 healthy controls). With a 0.5 test size the classification performance was 50.00 % with a MSE of 0.5 indicating poor performance of the classifier. The AUC is 0.67 indicating apoor sensitivity and specificity (Breiman, 2001; Geron, 2019). An f1 score of 0.86 was observed for the cancer group and of 0.50 for the healthy group (Figure 7C).

Decision trees are a supervised learning technique which use multiple algorithms to decide to split a node into two or more sub-nodes with tree-like diagrams to classify some target variable/data. The performance of the decision trees classifier is shown with a 0.5 test size on a randomly selected single frequency (at 1808.25 cm$^{-1}$). The classification accuracy was found to be 88.89%with a MSE of 0.111. The AUC is 0.94 indicating a high classification accuracy. An f1 score of 0.93 (i.e., precision of 1.00 and recall of 0.88) for cancer and 0.67 for healthy (i.e., precision of 0.50 and recall of 1.00) was observed. The entropy criterion (one of the learning parameters) was used for the tree classification. The results remained unchanged with baseline correction for the Decision Tree performance (Figure 7E).

The classification predictions by the SVM algorithm with a linear kernel is shown on the full range raw Raman spectra. SVM is a supervised ML algorithm which finds the optimal hyperplane that maximizes the margin between the data classes using gradient descent learning. Classification accuracy of 80.00% and MSE: 0.2 were observed. An f1 score of 0.75 for the cancer class (i.e., precision and recall of 0.75) and of 0.83 for the healthy class (i.e., precision and recall of 0.83) were obtained corresponding to an AUC of 0.79 (Figure 7G). The SVM classification accuracy on the raw spectra within the selected frequency range reported above and a 0.5 test size was found to be 88.89 % with a MSE of 0.111. The f1scores for the cancer and healthy group classification were 0.92 (precision of 0.86 and recall of 1.00) and 0.80 (precision of 1.00 and recall of 0.67), respectively (Figure 7I).

**SUPPLEMENTARY INFORMATION**

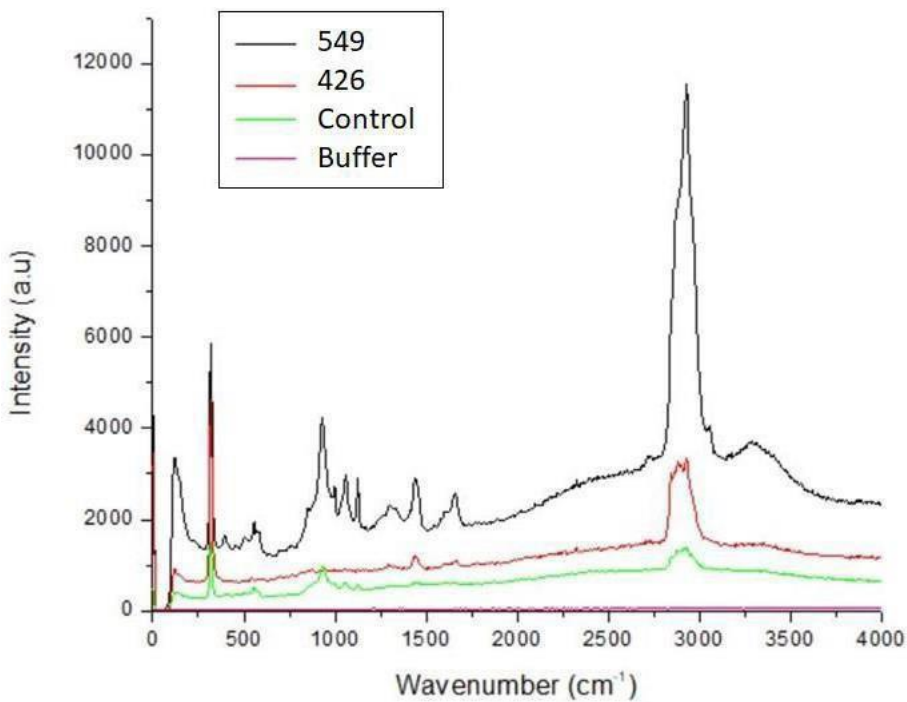

**FIGURE S1. RAMAN SPECTRA OF EVs SAMPLES I.** Representative raw Raman spectra of cancer and healthy control are shown as measured at a 514 nm excitation laser with 10% laser power, 10 accumulations for 30 seconds. EVs drops were air dried on the calcium fluoride slides prior to measurement acquisitions.

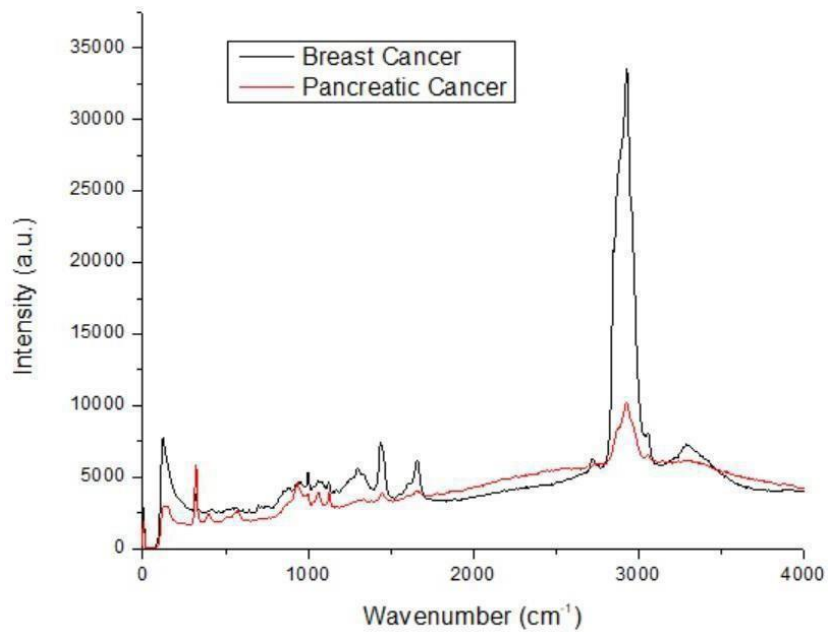

**FIGURE S2. RAMAN SPECTRA ON EVs SAMPLES II.** Single raw Raman spectra (without baseline correction) are shown to help visualize the spectral differences in the cancer EVs samples.

**PEAK FIT ASSIGNMENT**

Normalized and baseline-corrected spectra. Some curves show peaks which others do not. Only the most important peaks or peak ranges with significant features were discussed in the results section characterizing the differences between the patient groups (cancer vs. healthy).

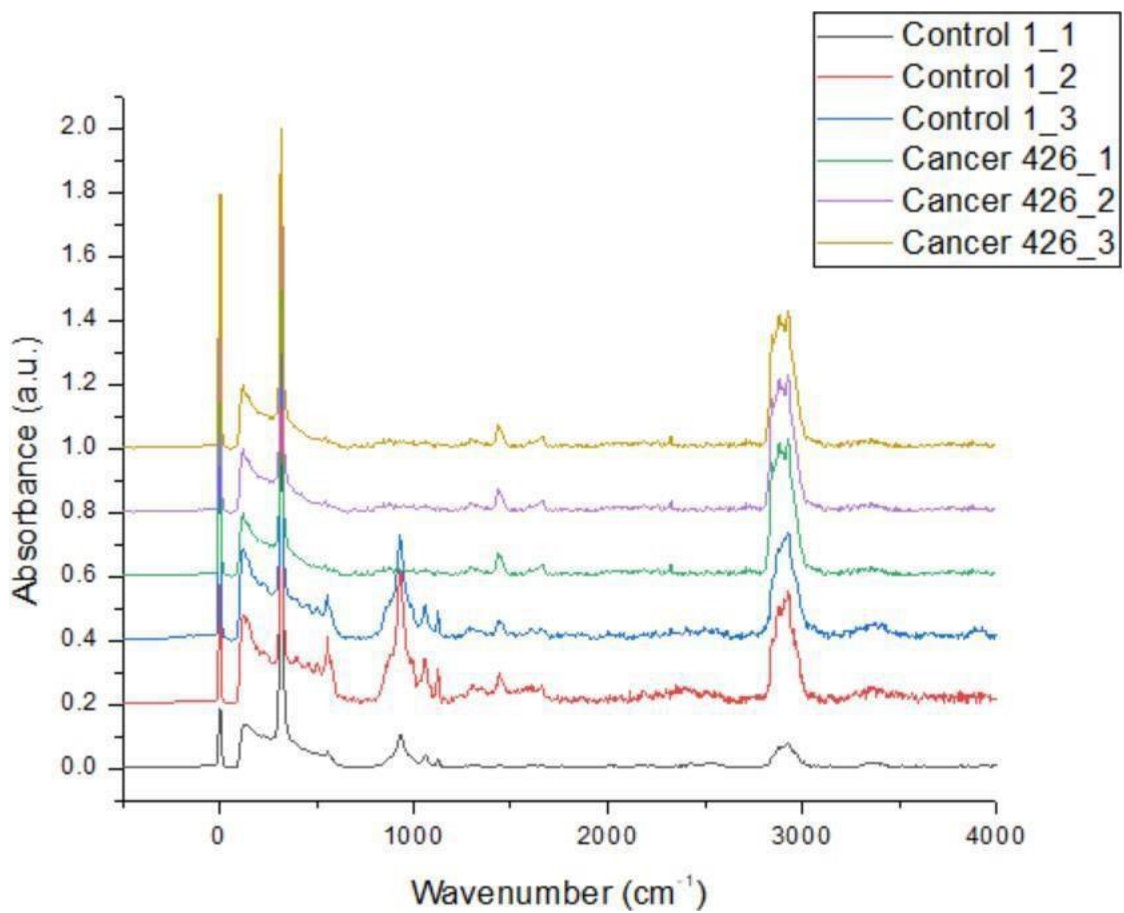

**FIGURE S3. RAMAN SPECTRA OF COLORECTAL CANCER SAMPLE 426.**

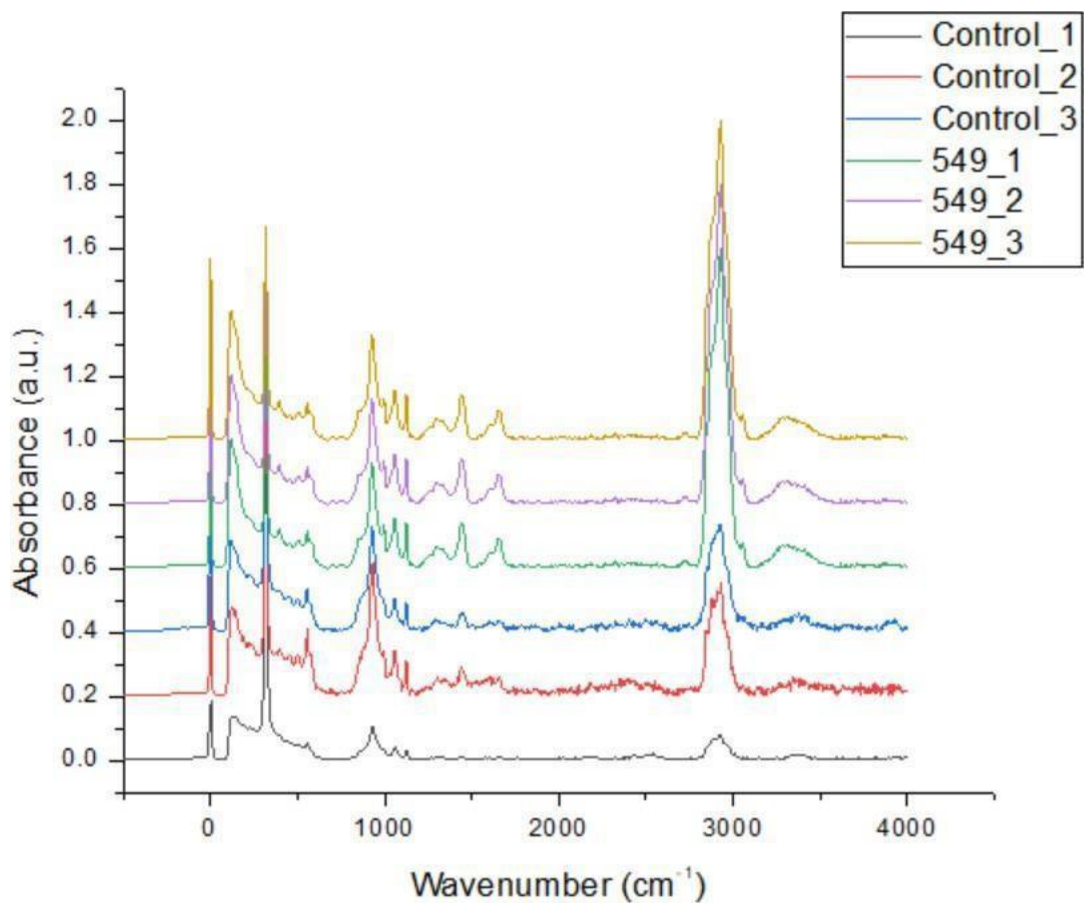

**FIGURE S4. RAMAN SPECTRA OF HEPATOCELLULAR CANCER SAMPLE 549.**

The following analysis applies to both the Raman spectra of 426 and 549. Three prominent peaks roughly at 930, 1062, and 1124 cm$^{-1}$. However, they are present in both groups: control and cancer spectra. The band found within the 2750-3040 cm$^{-1}$ range denotes the lipid band. Traditionally, the EV Raman spectra can be used to measure the spectroscopic protein-to-lipid (P/L) and nucleic acid-to-lipid (NA/L) ratio by their absorption bands. The P/L and NA/L ratio were estimated by dividing the relative intensity of amide I protein band (1600–1690 cm$^{-1}$) and nucleic acid band (720–800 cm$^{-1}$) by the lipid-band (2750–3040 cm$^{-1}$), respectively (Gualerzi et al., 2019). However, the use of ML algorithms as shown in our study provides a more robust quantitative method to detect patterns which distinguish healthy EVs from cancer-patient derived EVs.

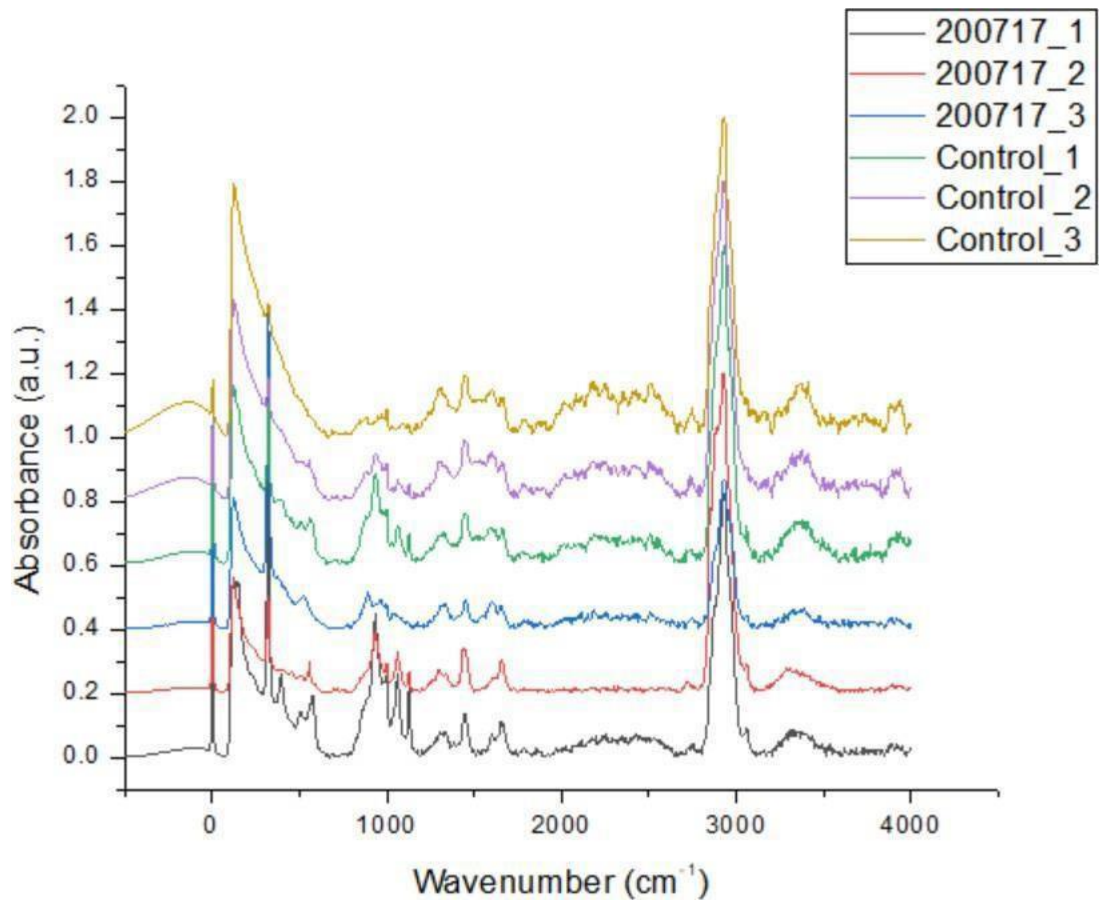

**FIGURE S5. RAMAN SPECTRA OF BREAST CANCER SAMPLE.**

Single peak at 2850-3010 cm-1 denotes the lipid band (C-H stretches). However, there appears to be/there is no obvious difference amidst the cancer vs. healthy groups. A peak at 720 cm$^{-1}$ may correspond to the presence of nucleic acids. No major differences are observed between the cancer and healthy groups suggesting further the premise that traditional approaches of peak-molecular bond assignment are insufficient in the study of such complex systems. The same analysis applies to pancreatic cancer (160517) spectra.

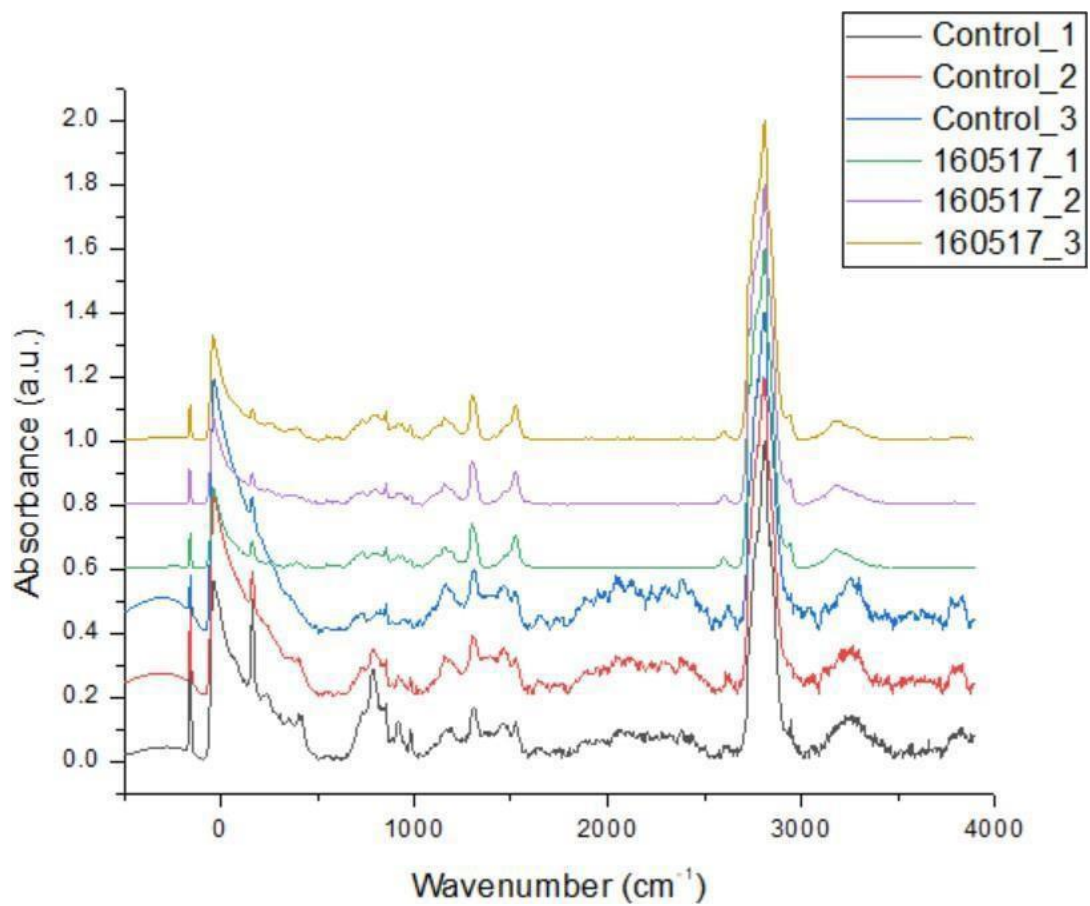

**FIGURE S6. RAMAN SPECTRA OF PANCREATIC CANCER SAMPLE.**

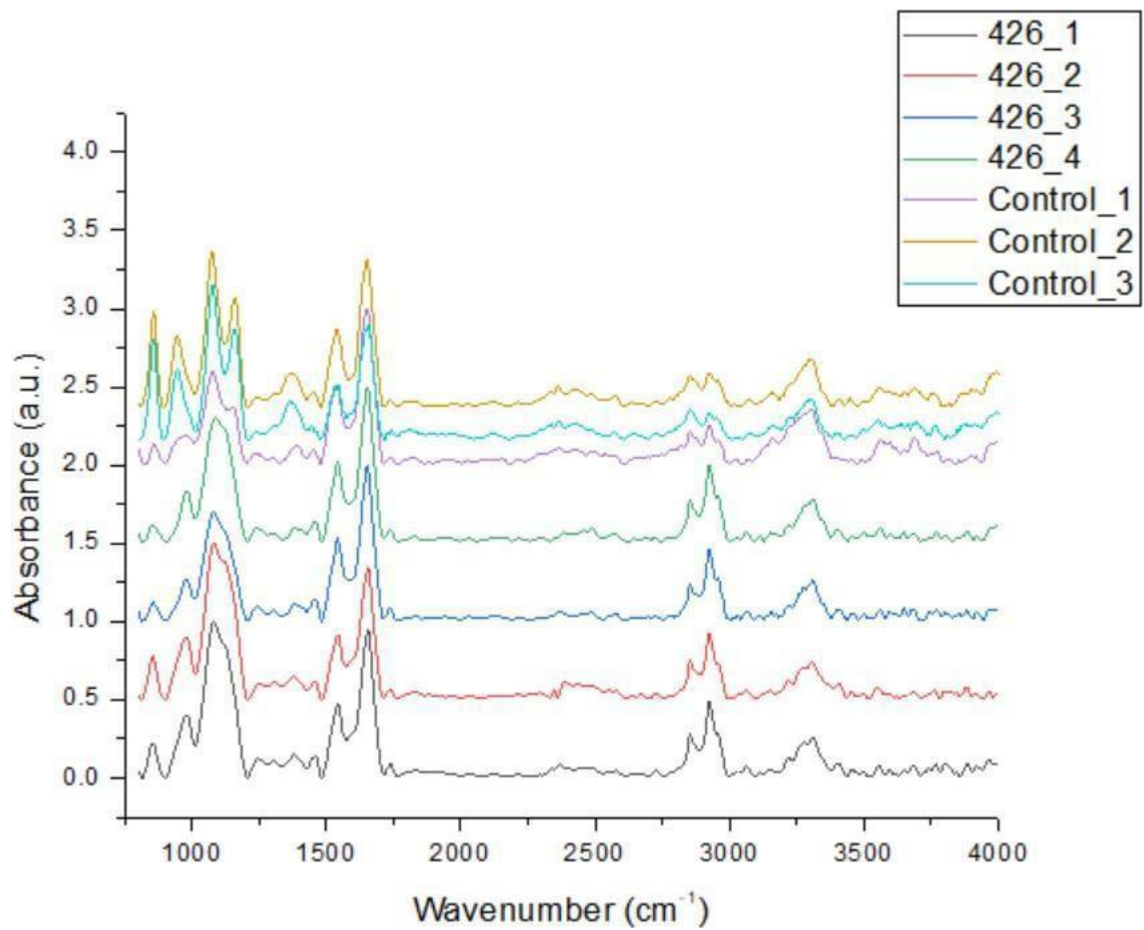

**FIGURE S7. FTIR SPECTRA OF 426.** Some characteristic peaks are found at: 1120 cm$^{-1}$ and at 3070 cm$^{-1}$ (the peaks are only present in the cancer spectra). There are additional peaks such as within the range of 1225-1450 cm$^{-1}$ and one at 1740 cm$^{-1}$ which were stronger (higher in absorption intensity) only in the cancer spectra.

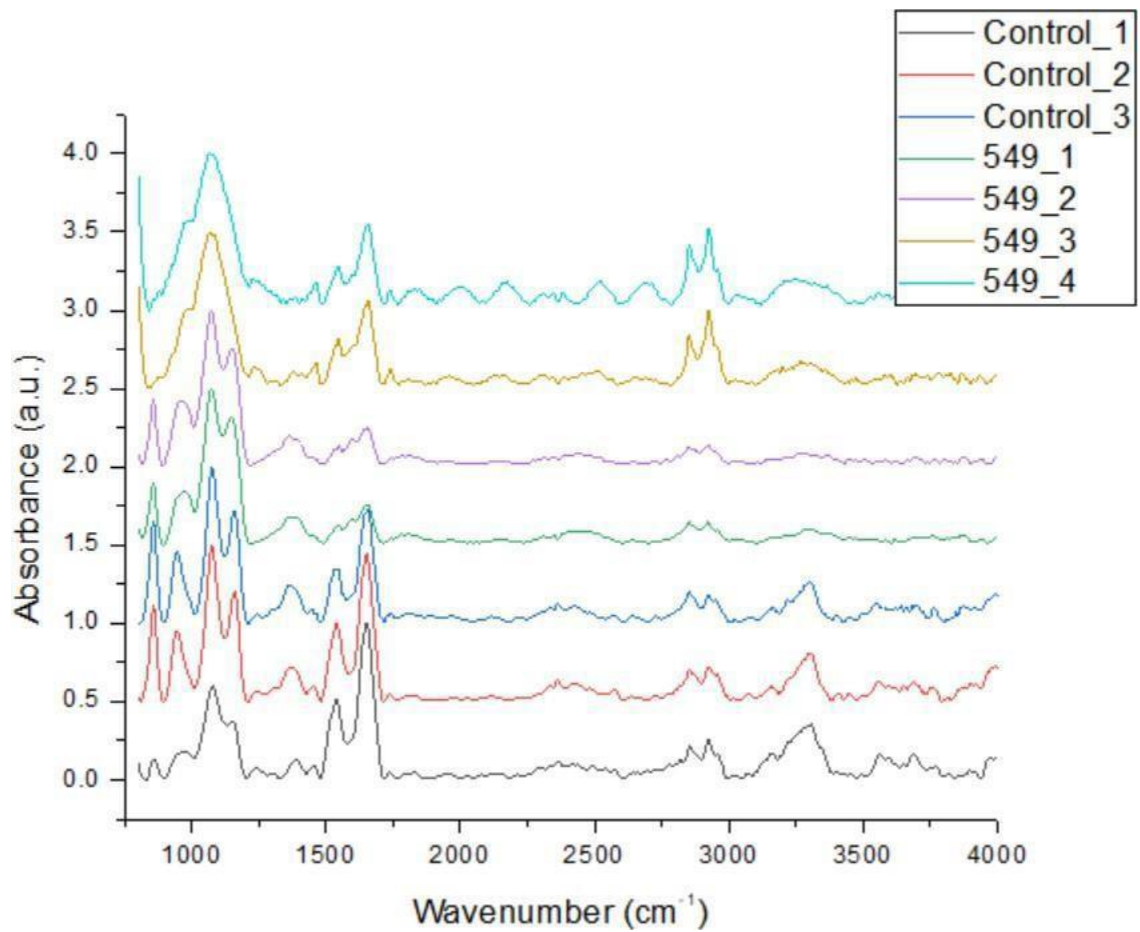

**FIGURE S8. FTIR SPECTRA OF 549.** The peak at 3070 cm$^{-1}$ is only present in cancer spectra. The peak at 980 cm$^{-1}$ is stronger in intensity in cancer spectra. Whether these distinct peaks are identified by the ML algorithms for the classification of cancer vs. healthy EVs remains unknown.

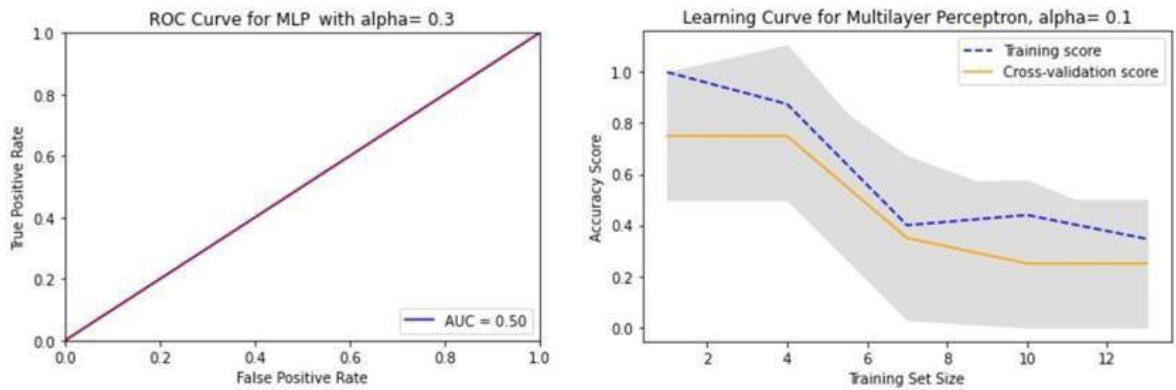

**FIGURE S9. ROC CURVE FOR MULTILAYER PERCEPTRON (MLP).** As discussed, the simple neural networkclassifier MLP had poorer performance compared to the classifiers discussed in the report above. Commonly researchers associate neural networks to optimal machine learning performance compared to simpler algorithms like SVM or Random Forests. However, these findings demonstrate this is not necessarily the case when dealing with smaller sample sizes. The performance metrics of AI systems in applications to interdisciplinary complex problems are highly model-dependent. Neural networks are in general, better suited for larger datasets, while smaller patient samples as in our study could benefit from simpler data mining algorithms. Furthermore, there remains a black box problem of how and which type of ML algorithms are better suited for a given dataset. Our findings clearly indicate decision trees/random forests and SVM are better in predicting vibrational spectra amidst other data mining algorithms such as the MLP or linear regression.

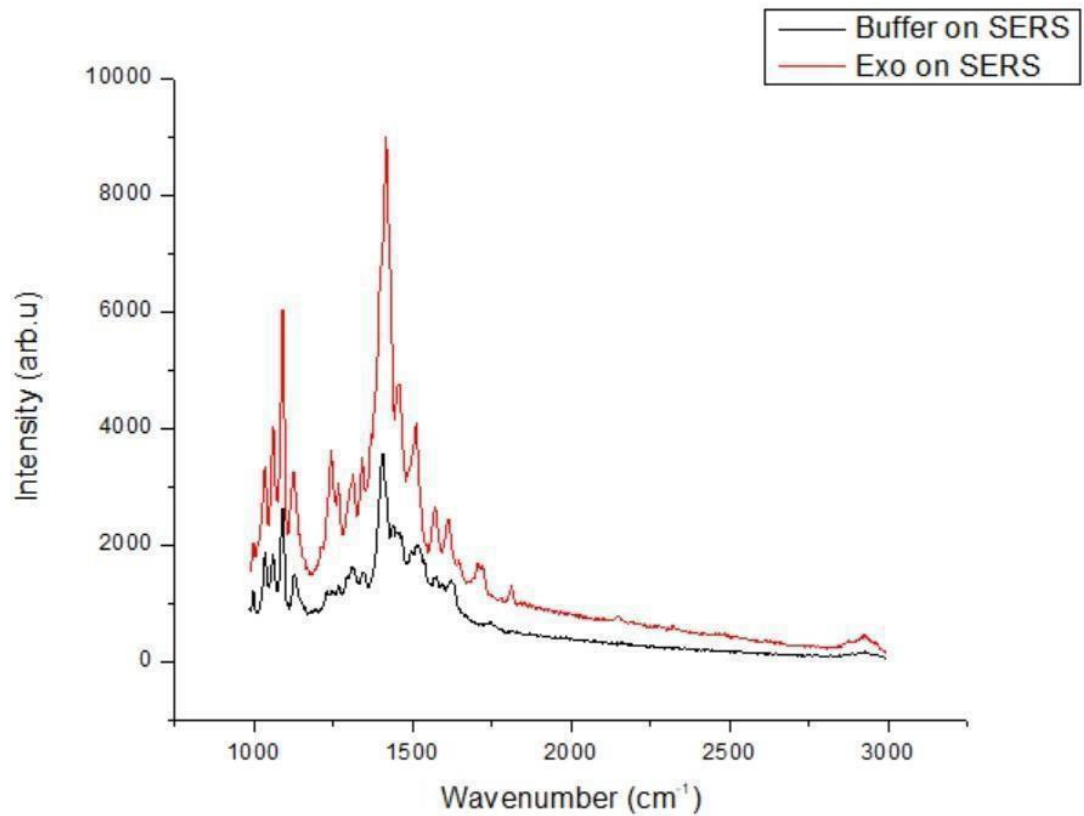

**FIGURE S10. SERS RAMAN SAMPLE SPECTRA.**

The raw SERS (surface enhanced Raman Spectroscopy) spectra on an EVs sample and PBS buffer at 50 X objective are shown. The SERS signal was acquired at 785 nm on an Au (gold) nanoparticle substrate (manufactured by Ocean Optics, Inc). Further runs were not followed since the gold-substrate consisted of a non-uniform paper. A purely homogeneous gold substrate is required to proceed with the SERS method. As shown in the data by Shin et al. (2018), the SERS spectra would provide detailed spectra and hence, allow the ML classifiers to better distinguish the cancer and healthy patient samples. The resolution of the Raman peaks are enhanced many-fold by the SERS technique in comparison to those reported in our study (on calcium fluoride slides). Despite these advantages the method with $CaF_2$ slides is cheaper and efficient as demonstrated by our findings. The demonstrated statistical results from the ML predictions indicate they are reasonably good in distinguishing the two patient groups, even when presented with a different set of heterogeneous (patient-derived) EVs samples.